%% file: 00_main.tex
\documentclass[sigconf]{acmart}

\pdfoutput=1

\usepackage{booktabs} 
\usepackage{framed}
\usepackage{hyperref}
\usepackage{gensymb}
\usepackage{subcaption}
\usepackage{tabularx}
\newcolumntype{Y}{>{\raggedright\arraybackslash}X}
\graphicspath{ {imgs/} }
\usepackage{pdfpages}

\usepackage[ruled]{algorithm2e}

\SetAlFnt{\small}
\SetAlCapFnt{\small}
\SetAlCapNameFnt{\small}
\SetAlCapHSkip{0pt}
\IncMargin{-\parindent}

\usepackage{color}

\newcommand{\systemname}{PrivacEye\xspace}

\def\plaintitle{\systemname: Privacy-Preserving Head-Mounted Eye Tracking Using Egocentric Scene Image and Eye Movement Features}

\usepackage{changes}
\colorlet{Changes@Color}{red}

\begin{document}
\sloppy

\title[\systemname: Privacy-Preserving Head-Mounted Eye Tracking]{\plaintitle}

\author{Julian Steil}
\affiliation{\institution{Max Planck Institute for Informatics\\ Saarland Informatics Campus}}
\email{jsteil@mpi-inf.mpg.de}

\author{Marion Koelle}
\affiliation{\institution{University of Oldenburg}}
\email{marion.koelle@uol.de}

\author{Wilko Heuten}
\affiliation{\institution{OFFIS - Institute for IT}}
\email{wilko.heuten@offis.de}

\author{Susanne Boll}
\affiliation{\institution{University of Oldenburg}}
\email{susanne.boll@uol.de}

\author{Andreas Bulling}
\affiliation{\institution{University of Stuttgart}}
\email{andreas.bulling@vis.uni-stuttgart.de}

\begin{abstract}
\input{01_abstract}
\vspace{-0.5cm}
\end{abstract}



\copyrightyear{2019}
\acmYear{2019}
\setcopyright{acmlicensed}
\acmConference[ETRA '19]{2019 Symposium on Eye Tracking Research and Applications}{June 25--28, 2019}{Denver, CO, USA}
\acmBooktitle{2019 Symposium on Eye Tracking Research and Applications (ETRA '19), June 25--28, 2019, Denver, CO, USA}
\acmPrice{15.00}
\acmDOI{10.1145/3314111.3319913}
\acmISBN{978-1-4503-6709-7/19/06}

\maketitle

\input{02_intro}
\input{03_relatedwork}
\input{04_approach}
\input{06_prototype}
\input{07_evaluation}
\input{08_discussion}
\input{09_conclusion}

\bibliographystyle{ACM-Reference-Format}
\bibliography{references}

\clearpage

\includepdf[pages=1-,pagecommand={}]{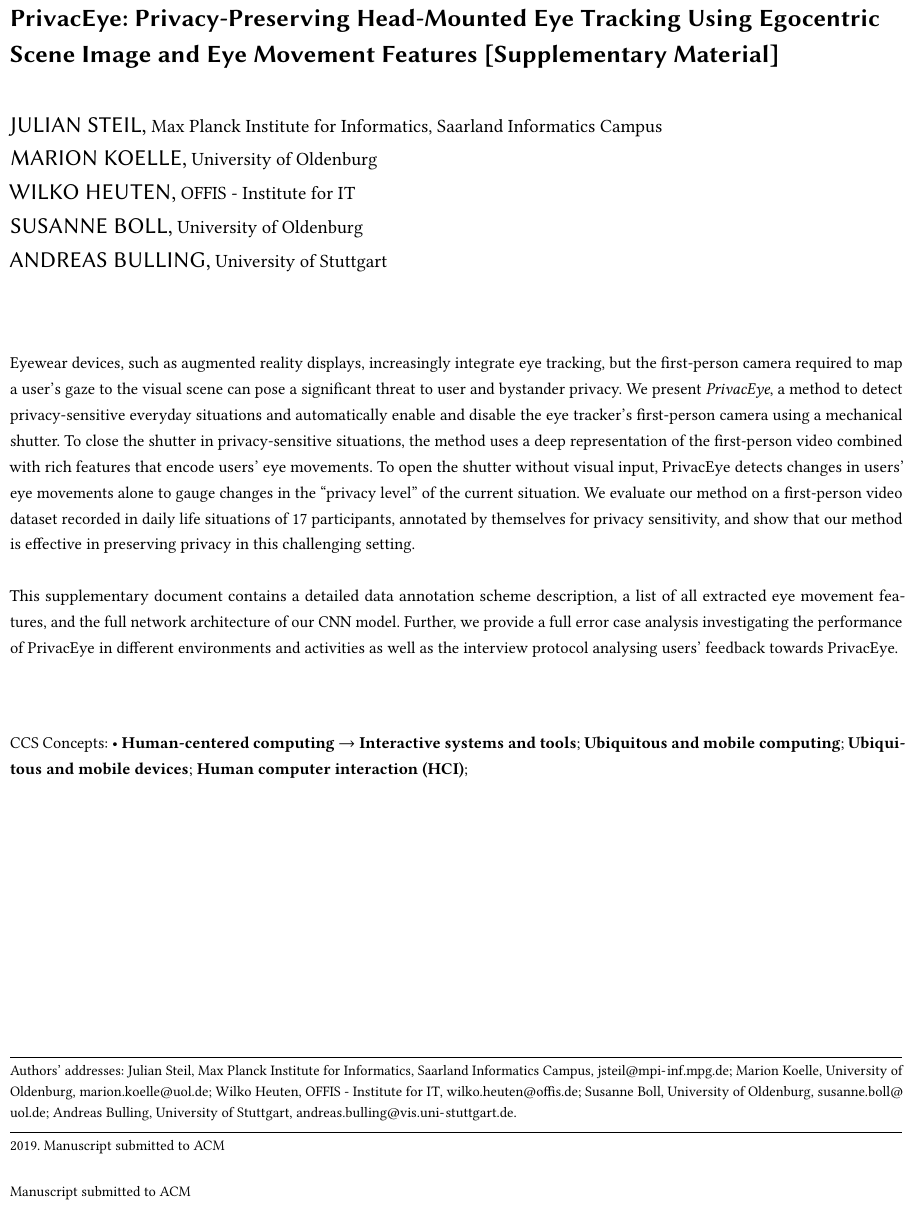}

\end{document}

%% file: 01_abstract.tex
Eyewear devices, such as augmented reality displays, increasingly integrate eye tracking, but the first-person camera required to map a user's gaze to the visual scene can pose a significant threat to user and bystander privacy. We present \textit{\systemname}, a method to detect privacy-sensitive everyday situations and automatically enable and disable the eye tracker's first-person camera using a mechanical shutter. To close the shutter in privacy-sensitive situations, the method uses a deep representation of the first-person video combined with rich features that encode users' eye movements. To open the shutter without visual input, \systemname detects changes in users' eye movements alone to gauge changes in the ``privacy level'' of the current situation. We evaluate our method on a first-person video dataset recorded in daily life situations of 17 participants, annotated by themselves for privacy sensitivity, and show that our method is effective in preserving privacy in this challenging setting.

%% file: 02_intro.tex
\vspace{-0.1cm}
\section{Introduction}

Eyewear devices, such as head-mounted displays or augmented reality glasses, have recently emerged as a new research platform in fields such as human-computer interaction, computer vision, or the behavioural and social sciences~\cite{bulling16_acmi}.
An ever-increasing number of these devices integrate eye tracking to analyse attention allocation~\cite{eriksen1985allocation,sugano2016aggregaze}, for computational user modelling~\cite{fischer2001user,itti2001computational}, or hands-free interaction~\cite{hansen2003command,vertegaal2003attentive}.
Head-mounted eye tracking typically requires two cameras: An eye camera that records a close-up video of the eye and a high-resolution first-person (scene) camera to map gaze estimates to the real-world scene~\cite{Kassner.2014}.
The scene camera poses a serious privacy risk 
as it may record sensitive personal information, such as login credentials, banking information, or text messages, as well as infringe on the privacy of bystanders~\cite{perez2017bystanders}.
Privacy risks intensify with the unobtrusive integration of eye tracking in ordinary glasses frames\mbox{~\cite{tonsen17_imwut}.}

\begin{figure} [tp]
\vspace{0.1cm}
  \centering
   \includegraphics[width=1\columnwidth]{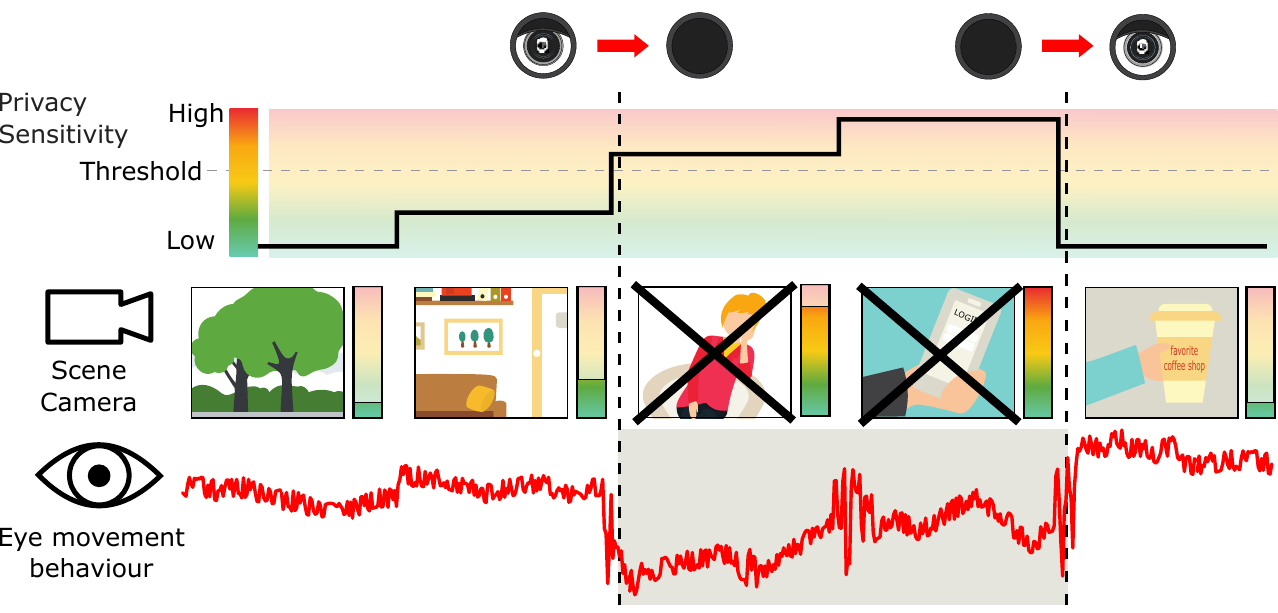}
   \vspace{-0.5cm}
  \caption{Our method uses a mechanical camera shutter (top) to preserve users' and bystanders' privacy with head-mounted eye trackers. Privacy-sensitive situations are detected by combining deep scene image and eye movement features (middle) while changes in eye movement behaviour alone trigger the reopening of the camera shutter (bottom).}
  \label{fig:teaser}
\end{figure}

In the area of first-person vision, prior work identified strategies of self-censorship~\cite{Koelle.2017} that, however, are prone to (human) misinterpretations and forgetfulness, or the accidental neglect of social norms and legal regulations.
In consequence, user experience and comfort are decreased and the user's mental and emotional load increases, while sensitive personal information can still be accidentally disclosed.
Other works therefore investigated alternative solutions, such as communicating a bystander's privacy preferences using short-range wireless radio~\cite{Aditya.2016}, visual markers~\cite{Schiff.2007}, or techniques to compromise recordings~\cite{Harvey.2012,Truong.2005}.
However, all of these methods require bystanders to take action themselves to protect their privacy.
None of these works addressed the problem at its source, i.e.\ the scene camera, nor did they offer a means to protect the privacy of both the wearer and potential bystanders.

To address this limitation, we propose~\textit{\systemname}, the first method for privacy-preserving head-mounted eye tracking (see Figure~\ref{fig:teaser}).
The key idea and core novelty of our method is to detect users' transitions into and out of privacy-sensitive everyday situations by leveraging both cameras available on these trackers.
If a privacy-sensitive situation is detected, the scene camera is occluded by a physical shutter.
Our design choice to use a non-spoofable physical shutter, which closes for some time and therefore provides feedback to bystanders, is substantiated by Koelle et al., who highlight an increased trustworthiness over LED lights on the camera or pure software solutions~\cite{koelle2018beyond}.
While this approach is secure and visible to bystanders, it prohibits visual input from the scene.
Thus, our method analyses changes in the users' eye movement behaviour alone to detect if they exit a privacy-sensitive situation and then reopens the camera shutter.
A naive, vision-only system could reopen the shutter at regular intervals, e.g.\ every 30 seconds, to detect whether the current situation is still privacy-sensitive.
However, this approach may negatively affect perceived reliability and increase mistrust in the system.
Thus, our eye-tracking approach promises significant advantages over a purely interval-based approach in terms of user experience and perceived trustworthiness.

Our approach is motivated by prior work 
that demonstrates
that eye movements are a rich source of information on a user's everyday activities~\cite{bulling11_pami,steil15_ubicomp}, social interactions and current environment~\cite{bulling13_chi}, or even a user's personality traits~\cite{hoppe18_fhns}.
In addition, prior work showed that perceived privacy sensitivity is related to a user's location and activity~\cite{Hoyle.2015}.
We therefore hypothesize that \textit{privacy sensitivity} transitively informs a user's \textit{eye movements}. 
We are the first to confirm this transitivity, which results as a reasoned deduction from 
prior work.

The specific contributions of this work are three-fold:
First, we present \textit{\systemname}, the first method that combines the analysis of egocentric scene image features with eye movement analysis to enable context-specific, privacy-preserving de-activation and re-activation of a head-mounted eye tracker's scene camera.
As such, we show a previously unconfirmed 
transitive relationship over the users' eye movements, their current activity and environment, as well as the perceived privacy sensitivity of the situation they are in.
Second, we evaluate our method on a dataset of real-world mobile interactions and eye movement data, fully annotated with locations, activities, and privacy sensitivity levels of 17 participants.
Third, we provide qualitative insights on the perceived social acceptability, trustworthiness, and desirability of \textit{\systemname}, based on semi-structured interviews, 
using
a fully functional 
prototype.

%% file: 03_relatedwork.tex
\section{Related Work}

Research on eye tracking privacy is sparse.
Thus, our work mostly relates to previous works on (1) privacy concerns with first-person cameras and (2) privacy enhancing methods for (wearable) cameras.

\subsection{Privacy Concerns - First-Person Cameras}
First-person cameras are well-suited for continuous and unobtrusive recordings,
which causes them to be perceived as 
unsettling by bystanders~\cite{Denning.2014}.
Both users' and bystanders' privacy concerns and attitudes towards head-mounted devices with integrated cameras were found to be affected by context, situation, usage intentions~\cite{Koelle.2015}, and user group~\cite{Profita.2016}.
Hoyle et al. showed that the presence and the number of people in a picture, specific objects (e.g., computer displays, ATM cards, physical documents), location, and activity affected whether lifeloggers deemed an image ``shareable''~\cite{Hoyle.2014}. %
They also highlighted the need for automatic privacy-preserving mechanisms to detect those elements, as individual sharing decisions are likely to be context-dependent and subjective.
Their results were partly confirmed by Price et al., who, however, found no significant differences in sharing when a screen was present~\cite{Price.2017}.
Chowdhury et al.\ found that whether lifelogging imagery is suitable for sharing is (in addition to content, scenario, and location) mainly determined by its sensitivity~\cite{Chowdhury.2016a}. %
Ferdous et al.\ proposed a set of guidelines that, among others, include semi-automatic procedures to determine the sensitivity of captured images according to user-provided preferences~\cite{Ferdous.2017}.
All highlight the privacy sensitivity of first-person recordings and the importance of protecting user and bystander privacy.

\subsection{Enhancing Privacy of First-Person Cameras}
To increase the privacy of first-person cameras for bystanders, researchers have suggested communicating their privacy preferences to nearby capture devices using wireless connections as well as mobile or wearable interfaces~\cite{Krombholz.2015}. 
Others have suggested preventing unauthorised recordings by compromising the recorded imagery, e.g., using infra-red light signals~\cite{Harvey.2010,yamada2013privacy} or disturbing face recognition~\cite{Harvey.2012}.
In contrast to our approach, these techniques all require the bystander to take action, which might be impractical due to costs and efforts~\cite{Denning.2014}.

A potential remedy are automatic, or semi-automatic approaches, such as
\textit{PlaceAvoider}, a technique that allows users to ``blacklist'' sensitive spaces, e.g.,
 bedroom or bathroom
~\cite{Templeman.2014}.
Similarly, \textit{ScreenAvoider} allowed users to control the disclosure of images of computer screens showing potentially \mbox{private} content~\cite{korayem2016enhancing}.
Erickson et al.\ proposed a method to identify security risks, such as ATMs, keyboards, and credit cards, in images captured by first-person wearable devices~\cite{ericksonneural}.
However, instead of assessing the whole scene in terms of privacy sensitivity, their systems only detected individual sensitive objects.
Raval et al.\ presented \textit{MarkIt}, a computer vision-based privacy marker framework that allowed users to use self-defined bounding boxes and hand-gestures to restrict visibility of content on two dimensional surfaces (e.g. white boards) or sensitive real-world objects~\cite{Raval.2014}. 
\textit{iPrivacy} automatically detects privacy-sensitive objects from social images users are willing to share using deep multi-task learning~\cite{yu2017iprivacy}. It warns the image owners what objects in the images need to be protected before sharing and 
recommends
privacy settings.

While all of these methods improved privacy, they either only did so post-hoc, i.e.\ after images had already been captured, or they required active user input.
In contrast, our approach aims to prevent potentially sensitive imagery from being recorded at all, automatically in the background, i.e.\ without engaging the user.
Unlike current computer vision based approaches that work in image space, e.g.\ by masking objects or faces~\cite{Raval.2014, Shu.2016, yamada2013privacy}, restricting access~\cite{korayem2016enhancing}, or deleting recorded images post-hoc~\cite{Templeman.2014}, we de-activate the camera completely using a mechanical shutter and also signal this to bystanders.
Our approach is the first to employ eye movement analysis for camera re-activation
that, unlike other sensing techniques (e.g., microphones, infra-red cameras), does not compromise the privacy of potential bystanders. 

%% file: 04_approach.tex
\section{Design Rationale}

\systemname's design rationale is based on user and bystander goals and expectations. In this section, we outline how \systemname's design contributes to avoiding erroneous disclosure of sensitive information, so-called misclosures (User Goal 1), and social friction (User Goal 2), and detail on three resultant design requirements.

\subsection{Goals and Expectations}
\textbf{Avoid Misclosure of Sensitive Data.} A user wearing smart glasses with an integrated camera would typically do so to make use of a particular functionality, e.g., visual navigation.
However, the device's ``always-on'' characteristic causes it to capture more than originally intended.
A navigation aid would require capturing  certain landmarks for tracking and localisation.
In addition, unintended imagery and potentially sensitive data is captured. 
Ideally, to prevent misclosures~\cite{Caine.2009}, sensitive data should not be captured.
However, requiring the user to constantly monitor her actions and environment for potential sensitive information (and then de-activate the camera manually) might increase the workload and cause stress. As users might be 
forgetful, misinterpret situations, or overlook privacy-sensitive items, automatic support from the system would be desirable from a user's perspective.

\noindent\textbf{Avoid Social Friction.} The smart glasses recording capabilities may cause social friction if they do not provide a clear indication whether the camera is on or off: Bystanders might even perceive device usage as a privacy threat when the camera is turned off~\cite{Koelle.2015, koelle2018beyond}.
In consequence, they feel uncomfortable around such devices~\cite{Bohn.2005, Denning.2014, Ens.2015, Koelle.2015}. 
Similarly, user experience is impaired when device users feel a need for justification as they could be accused of taking surreptitious pictures~\cite{Hakkila.2015, koelle2018beyond}.

\vspace{-0.1cm}
\subsection{Design Requirements}
As a consequence of these user goals there are three essential design requirements that \systemname addresses: (1) The user can make use of the camera-based functionality without the risk of misclosures or leakage of sensitive information.
(2) The system pro-actively reacts to the presence or absence of potentially privacy-sensitive situations and objects.
(3) The camera device communicates the recording status clearly to both user and bystander.

%% file: 06_prototype.tex
\section{\systemname Prototype}

\begin{figure}
    \centering
        \centering
        \includegraphics[width=1\columnwidth]{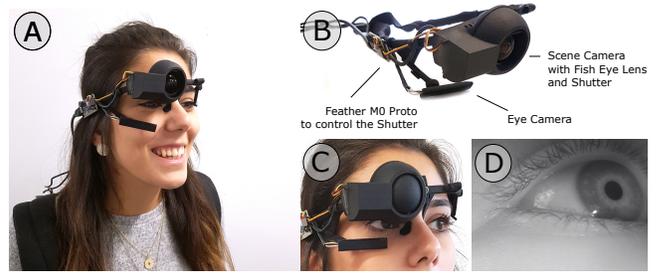} 
        \vspace{-0.6cm}
        \caption{\systemname prototype with labelled components (B) and worn by a user with a USB-connected laptop in a backpack (A). Detection of privacy-sensitive situations using computer vision closes the camera shutter (C), which is reopened based on a change in the privacy detected level in a user's eye movements (D).}
        \label{fig:setup}
\end{figure}

Our fully functional \systemname prototype, shown in Figure~\ref{fig:setup}, is based on the PUPIL head-mounted eye tracker~\cite{Kassner.2014} and features one 640$\times$480 pixel camera (the so-called ``eye camera'') that records the right eye from close proximity (30 fps), and a second camera (1280$\times$720 pixels, 24 fps) to record a user's environment (the so-called ``scene camera'').
The first-person camera is equipped with a fish eye lens with a 175$^\circ$ field of view and can be closed with a 
mechanical shutter. 
The shutter comprises a servo motor and a custom-made 3D-printed casing, including a mechanical lid to occlude the camera's lens. The motor and the lid are operated via a micro controller, namely a Feather M0 Proto.
Both cameras and 
the micro controller
were connected to a laptop via USB.
\systemname further consists of two main software components: (1) detection of privacy-sensitive situations to close the mechanical camera shutter and (2) detection of changes in user's eye movements that are likely to indicate suitable points in time for reopening the camera shutter.

\subsection{Detection of Privacy-Sensitive Situations}
The approaches for detecting privacy-sensitive situations we evaluated are (1) \textit{CNN-Direct}, (2) \textit{SVM-Eye}, and (3) \textit{SVM-Combined}.

\subsubsection{CNN-Direct}
Inspired by prior work on predicting privacy-sensitive pictures posted in social networks~\cite{orekondy17iccv}, we used a pre-trained GoogleNet, a 22-layer deep convolutional neural network~\cite{43022}.
We adapted the original GoogleNet model for our specific prediction task by adding two additional fully connected (FC) layers.
The first layer was used to reduce the feature dimensionality from 1024 to 68 and the second one, a Softmax layer, to calculate the prediction scores.
Output of our model was a score for each first-person image indicating whether the situation visible in that image was privacy-sensitive or not.
The cross-entropy loss was used to train the model.
The full network architecture is included in the supplementary material.

\subsubsection{SVM-Eye}
Given that eye movements are independent from the scene camera's shutter status, they can be used to (1) detect privacy-sensitive situations while the camera shutter is open and (2) detect changes in the subjective privacy level while the camera shutter is closed.
The goal of this second 
component
is to instead detect changes in a user's eye movements that are likely linked to changes in the privacy sensitivity of the current situation and thereby to keep the number of times the shutter is reopened as low as possible.
To detect privacy-sensitive situations and changes, we trained SVM classifiers (kernel=rbf, C=1) with characteristic eye movement features, which we extracted using only the eye camera video data.
We extracted a total of 52 eye movement features, covering fixations, saccades, blinks, and pupil diameter (see Table 2 in the supplementary material for a list and description of the features).
Similar to~\cite{bulling11_pami}, each saccade is encoded as a character forming words of length $n$ (wordbook).
We extracted these features using a sliding window of 30 seconds \mbox{(step size of 1 sec).}

\subsubsection{SVM-Combined}
A third approach for the detection of privacy-sensitive situations is a hybrid method. We trained SVM classifiers using the extracted eye movement features (52) and combined them with CNN features (68) from the scene image, which we extracted from the first fully connected layer of our trained CNN model,
 creating 120 feature large samples. With the concatenation of eye movement and scene features, we are able to extend the information from the two previous approaches during recording phases where the camera shutter is open.

%% file: 07_evaluation.tex
\section{Experiments}
We evaluated the different approaches on their own and in combination 
in a realistic temporal sequential analysis trained in a person-specific (leave-one-recording-out) and person-independent (leave-one-person-out) manner.
We assume that the camera shutter is open at start up. If no privacy-sensitive situation is detected, the camera shutter remains open and the current situation is rated ``non-sensitive'', otherwise, the camera shutter is closed and the current situation is rated ``privacy-sensitive''. 
Finally, we analysed error cases and investigated the performance of \systemname in different environments and activities.

\subsection{Dataset}

While an ever-increasing number of eye movement datasets have been published in recent years (see~\cite{steil15_ubicomp,bulling11_pami,bulling12_tap,hoppe18_fhns,Sugano_UIST15} for examples), none of them 
focused on privacy-related attributes.
We therefore make resource to a previously recorded dataset~\cite{steil2018forecasting}.
The dataset of Steil et al. contains more than 90 hours of data recorded continuously from 20 
participants (six females, aged 22-31) over more than four hours each. Participants were
students with different backgrounds and subjects with normal or corrected-to-normal vision.
During the recordings, participants roamed a university campus and performed their everyday activities, such as meeting people, eating, or working as they normally would on any day at the university.
To obtain some data from multiple, and thus also ``privacy-sensitive'', places on the university campus, participants were asked to not stay in one place for more than 30 minutes.
Participants were further asked to stop the recording after about one and a half hours so that the laptop's battery packs could be changed and the eye tracker re-calibrated.
This yielded three recordings of about 1.5 hours per participant.
Participants regularly interacted with a mobile phone provided to them and were also encouraged to use their own laptop, desktop computer, or music player if desired.
The dataset thus covers a rich set of representative real-world situations, including sensitive environments and tasks. 
The data collection \mbox{was performed with the} same equipment as shown in Figure~\ref{fig:setup} excluding the camera shutter.

\begin{figure} [t]
\vspace{-0.3cm}
  \centering
    \includegraphics[width=1\columnwidth]{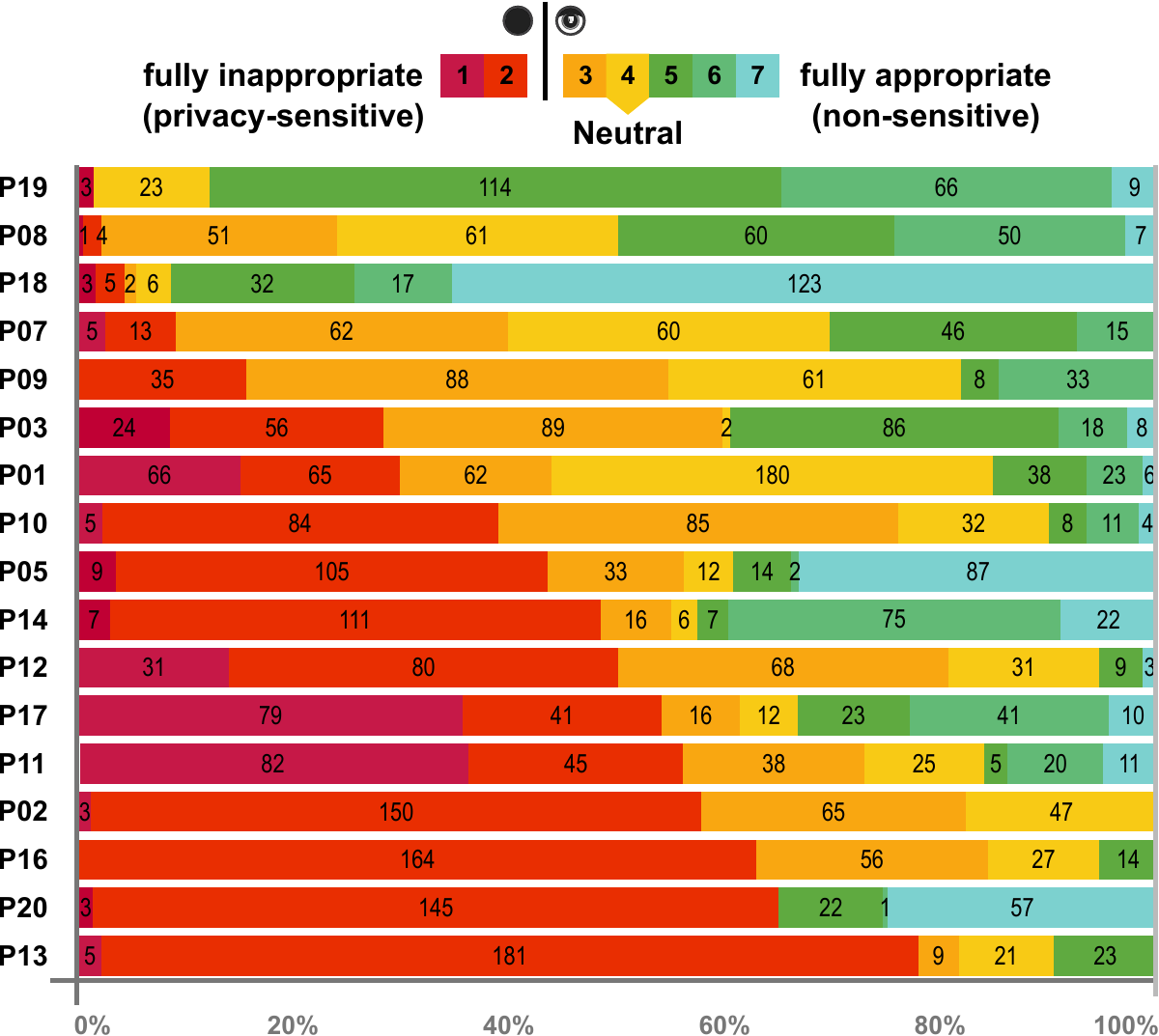}
    \vspace{-0.6cm}
  \caption{Privacy sensitivity levels rated on a 7-pt Likert scale from 1: fully inappropriate (i.e., privacy-sensitive) to 7: fully appropriate (i.e., non-sensitive). Distribution in labelled minutes/level per participant, sorted according to a ``cut-off'' between closed shutter (level 1-2) and open shutter (level 3-7). In practice, the ``cut-off'' level could be chosen according to individual ratings as measured by PAQ.}
  \label{fig:scoredistribution}
  \vspace{-0.2cm}
\end{figure}

\subsection{Data Annotation}
The dataset was fully annotated 
by the participants themselves
with
continuous annotations of location, activity, scene content, and subjective privacy sensitivity level.
17 out of the 20 participants finished the annotation 
of their own recording
resulting in about 70 hours of annotated video data.
They again gave informed consent and completed a questionnaire on demographics, social media experience and sharing behaviour~(based on Hoyle et al.~\cite{Hoyle.2014}), general privacy attitudes, as well as other-contingent privacy~\cite{Baruh.2014} and respect for bystander privacy~\cite{Price.2017}.
General privacy attitudes were assessed using the \textit{Privacy Attitudes Questionnaire}~(PAQ), a modified Westin Scale~\cite{Westin.2003} as used  
by~\cite{Caine.2009,Price.2017}.

\begin{figure*} [t]
\vspace{-0.3cm}
  \centering
  \begin{subfigure}{1\columnwidth}
    \centering
        \includegraphics[width=1\columnwidth]{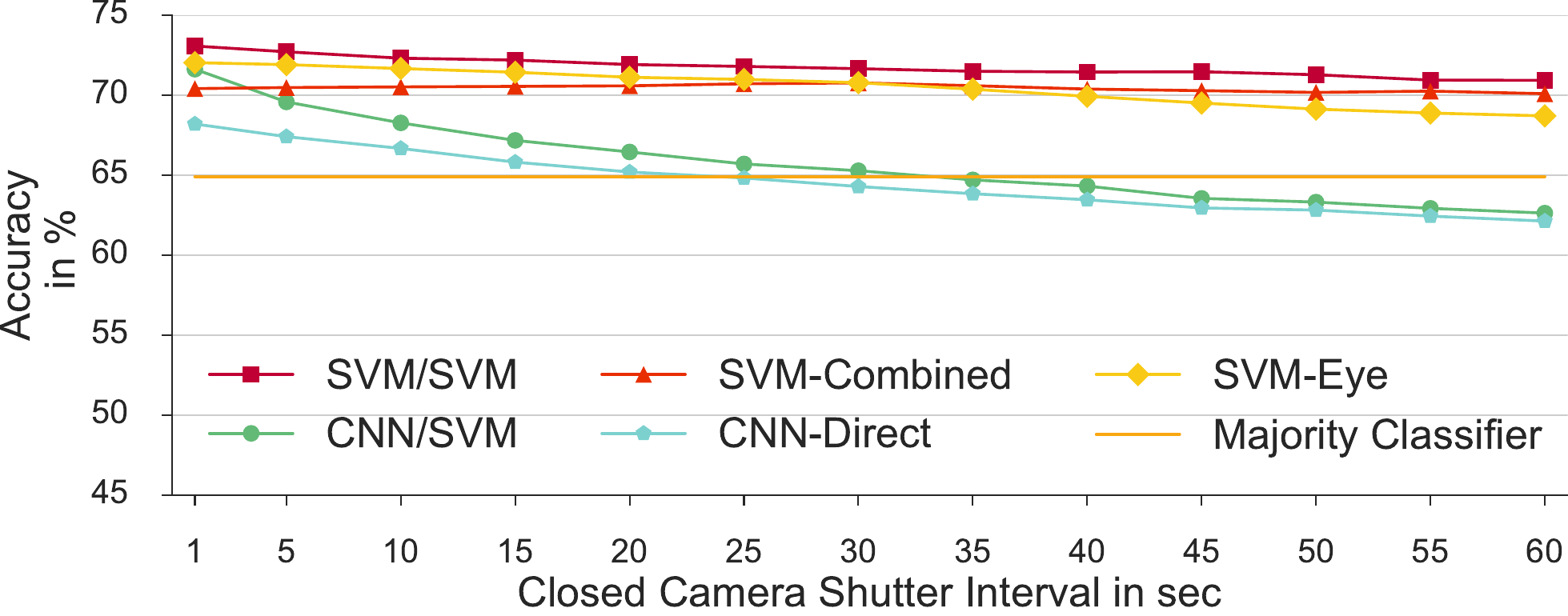}
        \caption{}
        \label{fig:Acc_pspec}
    \end{subfigure}
    \hfill
    \begin{subfigure}{1\columnwidth}
    \centering
        \includegraphics[width=1\columnwidth]{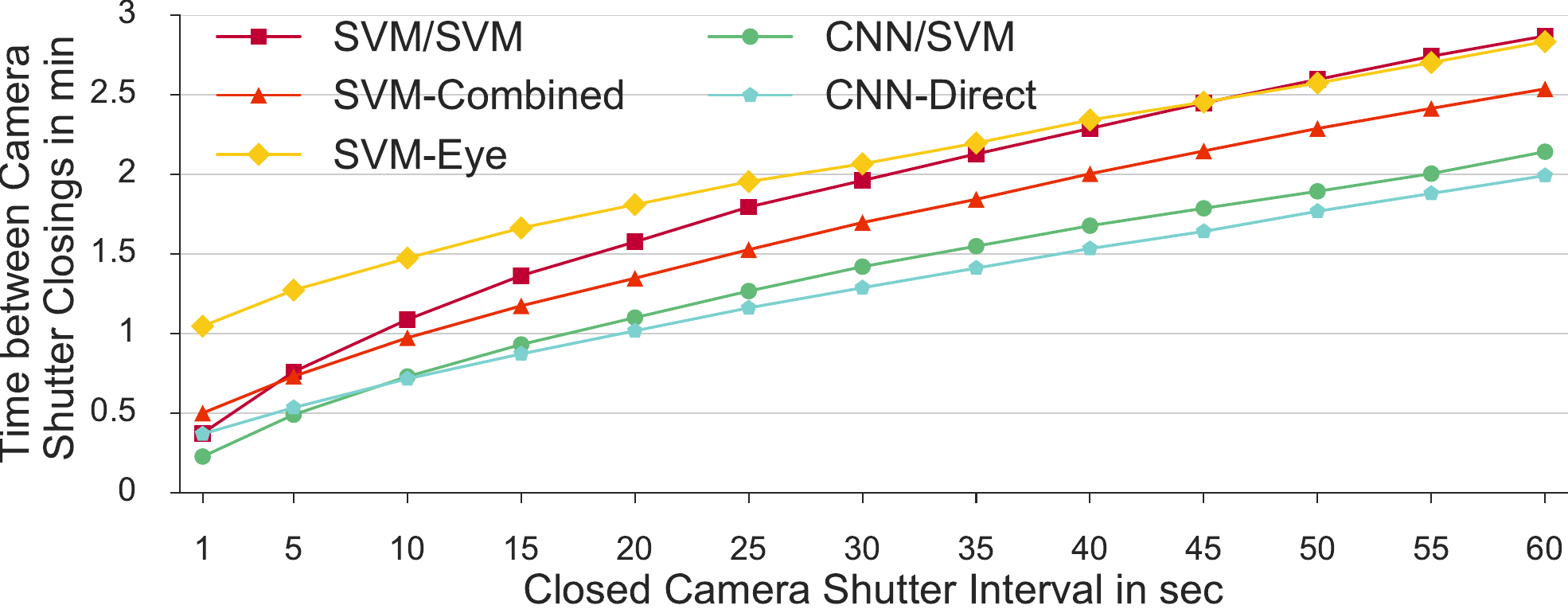}
        \caption{}
        \label{fig:Closing_pspec}
    \end{subfigure}
  \vspace{-0.3cm}
    \caption{Person-specific leave-one-recording-out evaluation showing the achieved accuracy (a) and the time between camera shutter closings (b) across different closed camera shutter intervals. 
  }
  \label{fig:PSPECEvaluation}
  \vspace{-0.2cm}
\end{figure*}

Annotations were performed using Advene~\cite{Aubert:2012:ATH:2361354.2361370}.
Participants were asked to annotate continuous video segments showing the same situation, environment, or activity.
They could also introduce new segments in case a privacy-relevant feature in the scene changed, e.g., when a participant switched to a sensitive app on the mobile phone.
Participants were asked to annotate each of these segments according to the annotation scheme (see supplementary material).
Privacy sensitivity was rated on a 7-point Likert scale 
ranging from 1 (fully inappropriate) to 7 (fully appropriate).
As we expected our participants to have difficulties understanding the concept of ``privacy sensitivity'', we rephrased it for the \mbox{annotation} to ``How appropriate is it that a camera is in the scene?''.
Figure~\ref{fig:scoredistribution} visualises the labelled privacy sensitivity levels for each participant.
Based on the latter distribution, we pooled ratings of 1 and 2 in the class ``privacy-sensitive'', and all others in the class ``non-sensitive''. 
A consumer system would provide the option to choose this ``cut-off''.
We will use these two classes for all evaluations and discussions that follow in order to show the effectiveness of our proof-of-concept system.
The dataset is available at \url{https://www.mpi-inf.mpg.de/MPIIPrivacEye/}.

\subsection{Sequential Analysis}

To evaluate \systemname, we applied the three proposed approaches separately as well as in combination in a realistic 
temporal
sequential analysis, evaluating the system as a whole within person-specific (leave-one-recording-out) and person-independent (leave-one-person-out) cross validation schemes.
Independent of CNN or SVM approaches, we first trained and then tested in a person-specific fashion.
That is, we trained on two of the three recordings of each participant and tested on the remaining one -- iteratively over all combinations and averaging the performance results in the end.
For the leave-one-person-out cross validation, we trained on the data of 16 participants and tested on the remaining one.
\textit{SVM-Eye} is the only one of the three proposed approaches that allows \systemname to be functional when no scene imagery is available, i.e., when the shutter is closed. Additionally, it can be applied when the shutter is open thus serving both software components of \systemname.
While the camera shutter is 
not closed, i.e., scene imagery is available,
\textit{CNN-Direct} or \textit{SVM-Combined} can be applied. 
To provide a comprehensive picture,
we then analysed the combinations \textit{CNN-Direct} + \textit{SVM-Eye} (\textit{CNN/SVM}) and \textit{SVM-Combined} + \textit{SVM-Eye} (\textit{SVM/SVM}).
The first approach is applied when the camera shutter is open and \textit{SVM-Eye} only when the shutter is closed. 
For the sake of completeness, we also evaluated \textit{SVM-Combined} and \textit{CNN-Direct} on the whole dataset.
However, these two methods represent hypothetical best-case scenarios in which eye and scene features are always available. 
As this is in practice not possible, they have to be viewed as an upper-bound baseline.
For evaluation purposes, we apply the proposed approaches within a step size of one second
in a sequential manner.
The previously predicted camera shutter position (open or close) decides which approach is applied for the prediction of the current state to achieve realistic results. 
We use $Accuracy=\frac{TP+TN}{TP+FP+TN+FN}$, where TP, FP, TN, and FN count sample-based true positives, false positives, true negatives, and false negatives, as performance indicator. 

\subsubsection{\textit{CNN-Direct}}
For training the CNN, which classifies a given scene image directly as privacy-sensitive or non-sensitive, we split the data from each participant into segments.
Each change in environment, activity, or the annotated privacy sensitivity level starts a new segment.
We used one random image per segment for training.

\subsubsection{\textit{SVM-Eye} and \textit{SVM-Combined}}
The SVM classifiers use only eye movement features (\textit{SVM-Eye}) or the combination of eye movement and CNN features (\textit{SVM-Combined}). We standardised the training data (zero mean, unit variance) for the person-specific and leave-one-person-out cross validation before training the classifiers, and used the same parameters for the test data.

\subsection{Results}

With potential usability implications in mind, we evaluate performance over a range of closed camera shutter intervals.
If a privacy-sensitive situation is detected from the \textit{CNN-Direct} or \textit{SVM-Combined} approach,
the camera shutter is kept closed for 
an interval 
between 1 and 60 seconds. If 
\textit{SVM-Eye} is applied and no privacy change is detected, the shutter remains closed.
In a practical application, users build
 more trust
 when the camera shutter remains closed, at least for a sufficient amount of time, to guarantee the protection of privacy-sensitive scene content when such a situation is detected~\cite{koelle2018beyond}. 
We also evaluated \textit{CNN-Direct} and \textit{SVM-Combined} 
on the whole recording 
as hypothetical best-case scenarios. However, comparing their performance against the combinations \textit{SVM/SVM} and \textit{CNN/SVM} illustrate the performance improvement using \textit{SVM-Eye} when the camera shutter is closed.

\begin{figure*} [t]
\vspace{-0.3cm}
  \centering
  \begin{subfigure}{1\columnwidth}
    \centering
        \includegraphics[width=1\columnwidth]{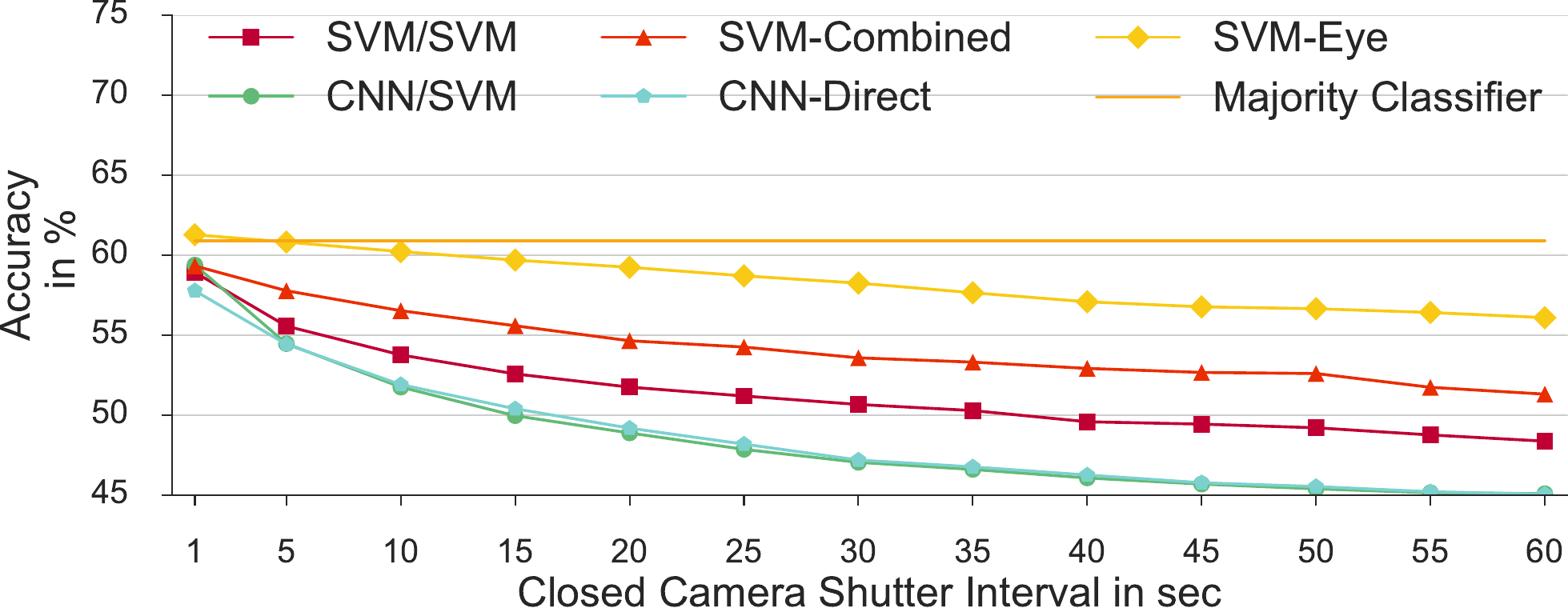}
        \vspace{-0.5cm}
        \caption{}
        \label{fig:Acc_lopo}
    \end{subfigure}
    \hfill
    \begin{subfigure}{1\columnwidth}
    \centering
        \includegraphics[width=1\columnwidth]{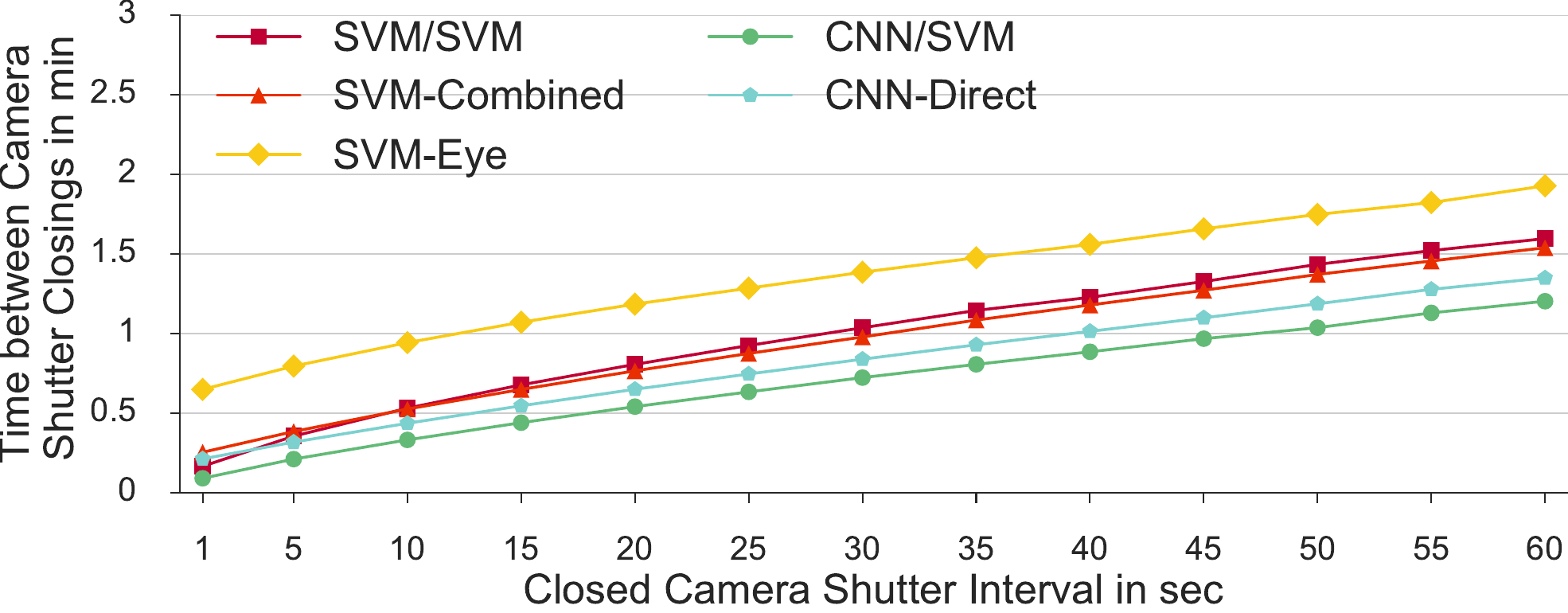}
        \vspace{-0.5cm}
        \caption{}
        \label{fig:Closing_lopo}
    \end{subfigure}
    \vspace{-0.3cm}
   \caption{Person-independent leave-one-person-out evaluation showing the accuracy results (a) and the time between closing the camera shutter (b) across different closed camera shutter intervals. 
  }
    \label{fig:LOPOEvaluation}
    \vspace{-0.2cm}
\end{figure*}

\subsubsection{Person-specific (leave-one-recording-out) evaluation}
Figure~\ref{fig:Acc_pspec} 
shows the person-specific accuracy performance of \systemname 
against increasing camera shutter closing time for two combinations
 \textit{CNN/SVM} and \textit{SVM/SVM}, and \textit{SVM-Eye}, which can be applied independent of the camera shutter status. Besides \textit{CNN-Direct} and \textit{SVM-Combined}, the majority class classifier 
serves as a baseline, 
predicting the majority class from the training set.
The results reveal that all trained approaches and combinations perform above the majority class classifier.
However, we can see that \textit{CNN-Direct} and its combination with \textit{SVM-Eye} 
(\textit{CNN/SVM})
perform below the other approaches 
and below the majority class classifier for longer closed camera shutter intervals.
\textit{SVM-Eye} and \textit{SVM-Combined} perform quite robustly, around 70\% accuracy, while \textit{SVM-Eye} performs better for shorter 
intervals and \textit{SVM-Combined} for longer \mbox{intervals.}
The interplay approach \textit{SVM/SVM}, which we would include in our prototype, exceeds 73\% with a closed camera shutter interval
 of one second and outperforms all other combinations in terms of accuracy in all other intervals.
One reason for the \mbox{performance} improvement of \textit{SVM/SVM} in comparison to its single components is that \textit{SVM-Combined} performs better for the detection of privacy-sensitive situations when the camera shutter is open while \textit{SVM-Eye} performs better for preserving privacy-sensitive situations so that the camera shutter remains closed.
Another aim of our proposed approach is the reduction of opening and closing events during a recording to strengthen
reliability and trustworthiness. 
A comparison of Figure~\ref{fig:Acc_pspec} and
 Figure~\ref{fig:Closing_pspec}
  renders a clear trade-off between accuracy performance and time between camera shutter closing instances. For very short camera shutter closing times the \textit{SVM-Eye} approach, which only relies on eye movement features from the eye camera, shows the best performance, whereas for longer camera shutter closing times, the combination \textit{SVM/SVM} shows better accuracy with a comparable amount of time between camera shutter closing instances. However, the current approaches are actually not able to reach the averaged ground truth of about 8.2 minutes between camera shutter closings.
 
\vspace{-0.1cm}
\subsubsection{Person-independent (leave-one-person-out) evaluation}
The more challenging task, which assumes that privacy-sensitivity could generalise over multiple participants, is given in the 
person-independent
leave-one-person-out cross validation of Figure~\ref{fig:Acc_lopo}.
Similar to the person-specific evaluation, \textit{CNN-Direct} and \textit{CNN/SVM} perform worse than the other approaches. 
Here, \textit{SVM-Eye} outperforms \textit{SVM-Combined} and \textit{SVM/SVM}. 
However, none of the approaches are able to outperform the majority classifier.
These results show that eye movement features generalise better over multiple participants to detect privacy-sensitive situations than scene image information. 
Comparing the number of minutes between camera shutter closing events of person-specific and leave-one-person-out in Figure~\ref{fig:Closing_pspec} and Figure~\ref{fig:Closing_lopo}, the person-specific approach outperforms the person-independent leave-one-person-out evaluation scheme for each approach. This shows that privacy sensitivity does not fully generalise, and consumer systems would require a person-specific calibration and online learning.

\vspace{-0.15cm}
\subsection{Error Case Analysis}

\begin{figure} [t]
\vspace{-0.2cm}
  \centering
      \includegraphics[width=1\columnwidth]{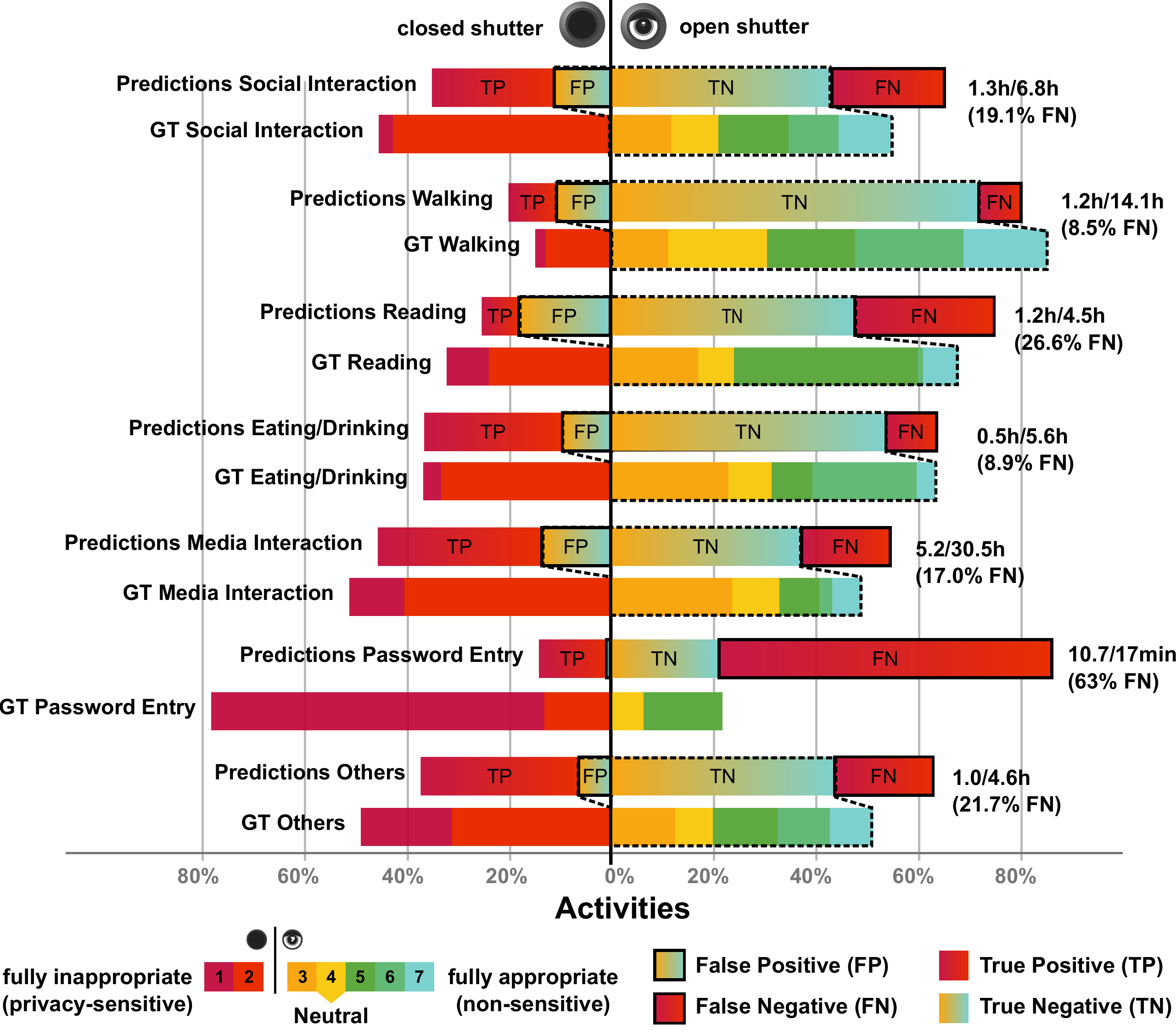}
  \vspace{-0.6cm}
  \caption{Error case analysis for different and activities showing the ``cut-off'' between closed shutter (left, \textit{privacy-sensitive}) and open shutter (right, \textit{non-sensitive}) with \systemname prediction and the corresponding ground truth (GT).  False positives (FP) are \textit{non-sensitive} but protected (closed shutter), false negatives (FN) are \textit{privacy-sensitive} but unprotected (open shutter).
  }
  \label{fig:PrivacyEvaluation}
  \vspace*{-1em}
\end{figure}

For \systemname, it is not only important to detect the privacy-sensitive situations (TP), but equally important to detect \mbox{non-} \mbox{sensitive} situations (TN), which are relevant 
to grant a good  
user experience.
Our results suggest that the combination \textit{SVM/SVM} performs best for the person-specific case. For this setting we carry out a detailed error case analysis of our system for the participants' different activities.
For the activities outlined in Figure~\ref{fig:PrivacyEvaluation}, \systemname works best while eating/drinking and in media interactions. Also, the results are promising for \mbox{detecting} social interactions. The performance for password entry, however, is still limited. Although the results show that it is possible to detect password entry, the amount of true negatives (TN) is high compared to other activities. This is likely caused by the dataset's under-representation of this activity, which characteristically lasts only a few seconds. Future work might be able to eliminate this by specifically training for password and PIN entry, which will enable the classifier to better distinguish between PIN entry and, e.g., reading.
In the supplementary material we provide an in-depth error case analysis to further investigate error cases in different environments.

\section{User Feedback}
Collecting initial subjective feedback during early stages of system development allows us to put research concepts in a broader context and helps to shape hypotheses for future quantitative user studies. %
In this section, we report on a set of semi-structured one-to-one interviews on the use of head-worn augmented reality displays in general, and our interaction design and prototype in particular. %
To obtain the user feedback, we recruited 12 new and distinct participants (six females), aged 21 to 31 years (M=24, SD=3) from the local student population.
They were enrolled in seven highly diverse majors, ranging from computer science and biology to special needs education. 
We decided to recruit students, given that we believe they and their peers are potential users of a future implementation of our prototype. 
We acknowledge that this sample, consisting of rather well educated young adults (with six of them having obtained a Bachelor's degree), is not representative for the general population.
Interviews lasted about half an hour and participants received a 5 Euro Amazon voucher. 
We provide a detailed interview protocol as part of the supplementary material. 
The semi-structured interviews were audio recorded and transcribed for later analysis. 
Subsequently, qualitative analysis was performed following inductive category development~\cite{Mayring.2014}.
Key motives and reoccurring themes were extracted and are presented in this section, where we link back to \systemname's design and discuss implications for future work.

\vspace{-0.1cm}
\subsection{User Views on Transparency}
Making it transparent (using the 3D-printed shutter), whether the camera was turned on or off, was valued by all participants. 
Seven participants found the integrated shutter increased perceived safety in contrast to current smart glasses; only few participants stated that they made no difference between the shutter and other visual feedback mechanisms, e.g. LEDs (n=2). 
Several participants noted that the physical coverage increased trustworthiness because it made the system more robust against hackers (\textit{concerns:hacking}, n=3) than LEDs. 
Concluding, the usage of physical occlusion could increase perceived safety and, thus, could be considered an option for future designs. 
Participants even noted that the usage of the shutter as reassuring as pasting up a laptop camera (\textit{laptop comparison}, n=4), which is common practice. 

\vspace{-0.1cm}
\subsection{User Views on Trustworthiness}
In contrast, participants also expressed technology scepticism, particularly that the system might secretly record audio (\textit{concerns:audio}, n=5) or malfunction (\textit{concerns:malfunction}, n=4). 
 With the increasing power of deep neural networks malfunctions, system failures, or inaccuracies will be addressable in the future, interaction designers will have to address this fear of ``being invisibly audio-recorded''. 
A lack of knowledge about eye tracking on both the user's and the bystander's side might even back this misconception. 
Therefore, future systems using eye tracking for context recognition will have to clearly communicate their modus operandi. 

\vspace{-0.1cm}
\subsection{Perceived Privacy of Eye Tracking}
The majority of participants claimed to have no privacy concerns about smart glasses with integrated eye tracking functionality: 
\textit{``I do see no threat to my privacy or the like from tracking my eye movements; this [the eye tracking] would rather be something which could offer a certain comfort.''}~(P11) Only two participants expressed \mbox{concerns} about their privacy, e.g., due to fearing eye-based emotion recognition~(P3). One was uncodeable. 
This underlines our assumption that eye tracking promises privacy-preserving and socially acceptable sensing in head-mounted augmented reality devices and, thus, should be further explored. 

\vspace{-0.1cm}
\subsection{Desired Level of Control}
Participants were encouraged to elaborate on whether the recording status should be user-controlled or system-controlled. 
P10 notes:
\textit{``I'd prefer if it was automatic, because if it is not automatic, then the wearer can forget to do that [de-activating the camera]. Or maybe he will say `Oh, I do not want to do that' and then [...] that leads to a conflict. So better is automatic, to avoid questions.''} Four other participants also preferred the camera to be solely controlled by the system (\textit{control:automatic}, n=4). 
Their preference is motivated by user forgetfulness (n=5), and potential non-compliance of users (in the bystander use case, n=1).
Only two participants expressed a preference for sole (\textit{control:manual}) control, due to an expected lack of system reliability, and technical feasibility. 
Two responses were uncodable. 
All other participants requested to implement manual confirmation of camera de-activation/re-activation or manual operation as alternative modes (\textit{control:mixed}, n=4), i.e., they like to feel in control. 
To meet these user expectations, future interaction designs would have to find an adequate mix of user control and automatic support through the system; 
for example, by enabling users to explicitly record sensitive information (e.g. in cases of emergency) or label seemingly non-sensitive situations ``confidential''.

%% file: 08_discussion.tex
\section{Discussion}

We discuss \systemname in light of the aforementioned design and user requirements and results of the technical evaluation.

\vspace{-0.1cm}
\subsection{Privacy Preserving Device Behaviour}
\textit{Design Requirements 1} and \textit{2} demand privacy-preserving device behaviour. With \systemname, we have presented a computer vision routine that analyses all imagery obtained from the scene camera, combined with eye movement features with regard to privacy sensitivity and, in case a situation requires protection, the ability to de-activate the scene camera and close the system's camera shutter. 
This approach prevents both accidental misclosure and malicious procurance (e.g. hacking) of sensitive data, as has been positively highlighted by our interview participants. 
However, closing the shutter comes at the cost of having the scene camera unavailable for sensing after it has been de-activated. 
\systemname solves this problem by using a second eye camera that allows us, in contrast to prior work, to locate all required sensing hardware on the user's side.
With \systemname we have provided 
proof-of-concept that context-dependent re-activation of a first-person scene camera is feasible using only eye movement data. 
Future work will be able to build upon these findings and further explore eye tracking as a sensor for privacy-enhancing technologies. 
Furthermore, our results provide 
first prove that there is indeed a transitive relationship 
over
privacy sensitivity 
and
a user's eye movements.

\vspace{-0.1cm}
\subsection{Defining Privacy Sensitivity}
Prior work indicates that the presence of a camera may be perceived appropriate or inappropriate depending on social context, location, or activity~\cite{ Hoyle.2014, Hoyle.2015, Price.2017}.
However, related work does, to the best of our knowledge, not provide any insights on eye tracking data in this context. 
For this reason, we run a dedicated data collection and ground truth annotation. 
Designing a practicable data collection experiment requires the overall time spent by a participant for data recording and annotation to be reduced to a reasonable amount. 
Hence, we made use of an already collected data set, and re-invited the participants only for the annotation task. 
While the pre-existing data set provided a rich diversity of privacy-sensitive locations and objects, including smart phone interaction, and realistically depicts everyday student life, it is most likely not applicable to other contexts, e.g., industrial work or medical scenarios. 

For \systemname, we rely on a 17-participant-large, ground truth annotated dataset with highly realistic training data.
Thus, the collected training data cannot be fully generalised, e.g., to other regions or age groups. 
On the plus side, however, this data already demonstrates that in a future real-world application, sensitivity ratings may vary largely between otherwise similar participants. 
This might also be affected by their (supposedly) highly individual definition of ``privacy''. 
Consequently, a future consumer system should be pre-trained and then adapted online, based on personalised retraining after user feedback. 
In addition, users should be enabled to select their individual ``cut-off'', i.e., the level from which a recording is blocked, which was set to ``2'' for \systemname. 
Future users of consumer devices might choose more rigorous or relaxed ``cut-off'' levels depending on their personal preference.
Initial user feedback also indicated that an interaction design that combines automatic, software-controlled de- and re-activation, with conscious control of the camera by the user, could be beneficial. 

\subsection{Eye Tracking for Privacy-Enhancement}
Eye tracking is advantageous for bystander privacy given that it only senses users and their eye movements. 
In contrast to, e.g., microphones or infra-red sensing, it senses a bystander and/or an environment only indirectly via the user's eye motion or reflections. 
Furthermore, eye tracking allows for implicit interaction and is non-invasive, and we expect it to become integrated into commercially available smart glasses in the near future. 
On the other hand, as noted by Liebling and Preibusch~\cite{liebling2014privacy,Preibusch.2014}, eye tracking data is a 
scare
resource, which can be used to identify
user attributes like age, gender, health, or user's current task.
For this reason, the collection and use of eye tracking data could be perceived as a potential threat to user privacy. 
However, our interviews showed that eye tracking was not perceived as problematic by a large majority of our participants. 
Nevertheless, eye tracking data must be protected by appropriate privacy policies and data hygiene.

To use our proposed hardware prototype in a real-world scenario, data sampling and analysis need to run on a mobile phone. The CNN feature extraction is 
currently
the biggest computational bottleneck,
but could be implemented in hardware to allow for real-time operation (c.f., Qualcom's Snapdragon 845).
Further, we believe that a 
consumer system
should provide an accuracy >90\% which could be achieved using additional sensors such as GPS or inertial tracking. However, presenting the first approach for automatic de- and re-activation of a first-person camera that achieves $\sim$73\% with competitive performance to \textit{ScreenAvoider} (54.2 - 77.7\%)~\cite{Korayem.2014} and \textit{iPrivacy} ($\sim$75\%)~\cite{yu2017iprivacy}, which are restricted to scene content protection and post-hoc privacy protection, we 
provide a solid
basis for follow up work.
We note that
a generalized person-independent model for privacy sensitivity protection is desirable. For this work only the participants themselves labelled their own data. Aggregated labels of multiple annotators would result in a more consistent and generalizable ``consensus'' model and improve test accuracy, but would dilute the measure of perceived privacy sensitivity, which is highly subjective~\cite{Price.2017}. Specifically, similar activities and environments were judged differently by the individual participants,
as seen in Figure~\ref{fig:scoredistribution}.
 The availability of this information is a core contribution of our dataset. 

\subsection{Communicating Privacy Protection}
The interaction design of \systemname tackles \textit{Design Requirement 3} using a non-transparent shutter.
Ens et al.~\cite{Ens.2015} reported that the majority of their participants expected to feel more comfortable around a wearable camera device if it clearly indicated to be turned on or off. 
Hence, our proposed interaction design aims to improve a bystander's awareness of the recording status by employing an \textit{eye metaphor}. 
Our prototype implements the ``eye lid'' as a retractable shutter made from non-transparent material: open when the camera is active, closed when the camera is 
inactive.
Thus, the metaphor mimics ``being watched'' by the camera.
The ``eye lid'' shutter ensures that bystanders can comprehend the recording status without prior knowledge, as eye metaphors have been widely employed for interaction design, e.g., to distinguish visibility or information disclosure~\cite{Motti.2016, Pousman.2004, Schlegel.2011} or to signal user attention~\cite{Chan.2017}.
Furthermore, in contrast to visual status indicators, such as point lights (LEDs), physical occlusion is non-spoofable (c.f.,~\cite{Denning.2014, Portnoff.2015}).
This concept has been highly appreciated during our interviews, which is why we would recommend adopting it for future hardware designs.

%% file: 09_conclusion.tex
\section{Conclusion}

In this work, we have proposed \systemname, a method that combines first-person computer vision with eye movement analysis to enable context-specific, privacy-preserving de-activation and re-activation of a head-mounted eye tracker's scene camera.
We have evaluated our method quantitatively on a 17-participant dataset of fully annotated everyday behaviour as well as qualitatively, by collecting subjective user feedback from 12 potential future users.
To the best of our knowledge, our method is the first of its kind and prevents potentially sensitive imagery from being recorded at all, without the need for active user input.
As such, we believe the method opens up a new and promising direction for future work in head-mounted eye tracking, the importance of which will only increase with further miniaturisation and integration of eye tracking in head-worn devices or even in normal glasses frames.

\begin{acks}
This work was funded, in part, by a \grantsponsor{}{JST CREST}{} research grant under Grant No.:~\grantnum{1}{JPMJCR14E1}, Japan.
\end{acks}

%% file: 00_main.bbl

\begin{thebibliography}{58}


\ifx \showCODEN    \undefined \def \showCODEN     #1{\unskip}     \fi
\ifx \showDOI      \undefined \def \showDOI       #1{#1}\fi
\ifx \showISBNx    \undefined \def \showISBNx     #1{\unskip}     \fi
\ifx \showISBNxiii \undefined \def \showISBNxiii  #1{\unskip}     \fi
\ifx \showISSN     \undefined \def \showISSN      #1{\unskip}     \fi
\ifx \showLCCN     \undefined \def \showLCCN      #1{\unskip}     \fi
\ifx \shownote     \undefined \def \shownote      #1{#1}          \fi
\ifx \showarticletitle \undefined \def \showarticletitle #1{#1}   \fi
\ifx \showURL      \undefined \def \showURL       {\relax}        \fi
\providecommand\bibfield[2]{#2}
\providecommand\bibinfo[2]{#2}
\providecommand\natexlab[1]{#1}
\providecommand\showeprint[2][]{arXiv:#2}

\bibitem[\protect\citeauthoryear{Aditya, Sen, Druschel, Joon~Oh, Benenson,
  Fritz, Schiele, Bhattacharjee, and Wu}{Aditya et~al\mbox{.}}{2016}]%
        {Aditya.2016}
\bibfield{author}{\bibinfo{person}{Paarijaat Aditya},
  \bibinfo{person}{Rijurekha Sen}, \bibinfo{person}{Peter Druschel},
  \bibinfo{person}{Seong Joon~Oh}, \bibinfo{person}{Rodrigo Benenson},
  \bibinfo{person}{Mario Fritz}, \bibinfo{person}{Bernt Schiele},
  \bibinfo{person}{Bobby Bhattacharjee}, {and} \bibinfo{person}{Tong~Tong Wu}.}
  \bibinfo{year}{2016}\natexlab{}.
\newblock \showarticletitle{I-pic: A platform for Privacy-compliant Image
  Capture}. In \bibinfo{booktitle}{\emph{Annual International Conference on
  Mobile Systems, Applications, and Services (MobiSys)}}. ACM,
  \bibinfo{pages}{235--248}.
\newblock
\urldef\tempurl%
\url{https://doi.org/10.1145/2906388.2906412}
\showDOI{\tempurl}


\bibitem[\protect\citeauthoryear{Aubert, Pri{\'e}, and Schmitt}{Aubert
  et~al\mbox{.}}{2012}]%
        {Aubert:2012:ATH:2361354.2361370}
\bibfield{author}{\bibinfo{person}{Olivier Aubert}, \bibinfo{person}{Yannick
  Pri{\'e}}, {and} \bibinfo{person}{Daniel Schmitt}.}
  \bibinfo{year}{2012}\natexlab{}.
\newblock \showarticletitle{Advene As a Tailorable Hypervideo Authoring Tool: A
  Case Study}. In \bibinfo{booktitle}{\emph{Proceedings of the 2012 ACM
  Symposium on Document Engineering}} \emph{(\bibinfo{series}{DocEng '12})}.
  \bibinfo{publisher}{ACM}, \bibinfo{address}{New York, NY, USA},
  \bibinfo{pages}{79--82}.
\newblock
\showISBNx{978-1-4503-1116-8}
\urldef\tempurl%
\url{https://doi.org/10.1145/2361354.2361370}
\showDOI{\tempurl}


\bibitem[\protect\citeauthoryear{Baruh and Cemalc{\i}lar}{Baruh and
  Cemalc{\i}lar}{2014}]%
        {Baruh.2014}
\bibfield{author}{\bibinfo{person}{Lemi Baruh} {and} \bibinfo{person}{Zeynep
  Cemalc{\i}lar}.} \bibinfo{year}{2014}\natexlab{}.
\newblock \showarticletitle{It is more than personal: Development and
  validation of a multidimensional privacy orientation scale}.
\newblock \bibinfo{journal}{\emph{Personality and Individual Differences}}
  \bibinfo{volume}{70} (\bibinfo{year}{2014}), \bibinfo{pages}{165--170}.
\newblock
\urldef\tempurl%
\url{https://doi.org/DOI: 10.1016/j.paid.2014.06.042}
\showDOI{\tempurl}


\bibitem[\protect\citeauthoryear{Bohn, Coroam{\u{a}}, Langheinrich, Mattern,
  and Rohs}{Bohn et~al\mbox{.}}{2005}]%
        {Bohn.2005}
\bibfield{author}{\bibinfo{person}{J{\"u}rgen Bohn}, \bibinfo{person}{Vlad
  Coroam{\u{a}}}, \bibinfo{person}{Marc Langheinrich},
  \bibinfo{person}{Friedemann Mattern}, {and} \bibinfo{person}{Michael Rohs}.}
  \bibinfo{year}{2005}\natexlab{}.
\newblock \showarticletitle{Social, economic, and ethical implications of
  ambient intelligence and ubiquitous computing}.
\newblock In \bibinfo{booktitle}{\emph{Ambient Intelligence}}.
  \bibinfo{publisher}{Springer}, \bibinfo{pages}{5--29}.
\newblock
\urldef\tempurl%
\url{https://doi.org/10.1007/3-540-27139-2_2}
\showDOI{\tempurl}


\bibitem[\protect\citeauthoryear{Bulling and Kunze}{Bulling and Kunze}{2016}]%
        {bulling16_acmi}
\bibfield{author}{\bibinfo{person}{Andreas Bulling} {and} \bibinfo{person}{Kai
  Kunze}.} \bibinfo{year}{2016}\natexlab{}.
\newblock \showarticletitle{Eyewear Computers for Human-Computer Interaction}.
\newblock \bibinfo{journal}{\emph{ACM Interactions}} \bibinfo{volume}{23},
  \bibinfo{number}{3} (\bibinfo{year}{2016}), \bibinfo{pages}{70--73}.
\newblock
\showISBNx{1072-5520}
\urldef\tempurl%
\url{https://doi.org/10.1145/2912886}
\showDOI{\tempurl}


\bibitem[\protect\citeauthoryear{Bulling, Ward, and Gellersen}{Bulling
  et~al\mbox{.}}{2012}]%
        {bulling12_tap}
\bibfield{author}{\bibinfo{person}{Andreas Bulling}, \bibinfo{person}{Jamie~A.
  Ward}, {and} \bibinfo{person}{Hans Gellersen}.}
  \bibinfo{year}{2012}\natexlab{}.
\newblock \showarticletitle{Multimodal Recognition of Reading Activity in
  Transit Using Body-Worn Sensors}.
\newblock \bibinfo{journal}{\emph{ACM Transactions on Applied Perception}}
  \bibinfo{volume}{9}, \bibinfo{number}{1} (\bibinfo{year}{2012}),
  \bibinfo{pages}{2:1--2:21}.
\newblock
\urldef\tempurl%
\url{https://doi.org/10.1145/2134203.2134205}
\showDOI{\tempurl}


\bibitem[\protect\citeauthoryear{Bulling, Ward, Gellersen, and
  Tr{\"{o}}ster}{Bulling et~al\mbox{.}}{2011}]%
        {bulling11_pami}
\bibfield{author}{\bibinfo{person}{Andreas Bulling}, \bibinfo{person}{Jamie~A.
  Ward}, \bibinfo{person}{Hans Gellersen}, {and} \bibinfo{person}{Gerhard
  Tr{\"{o}}ster}.} \bibinfo{year}{2011}\natexlab{}.
\newblock \showarticletitle{Eye {M}ovement {A}nalysis for {A}ctivity
  {R}ecognition {U}sing {E}lectrooculography}.
\newblock \bibinfo{journal}{\emph{IEEE Transactions on Pattern Analysis and
  Machine Intelligence}} \bibinfo{volume}{33}, \bibinfo{number}{4}
  (\bibinfo{date}{April} \bibinfo{year}{2011}), \bibinfo{pages}{741--753}.
\newblock
\urldef\tempurl%
\url{https://doi.org/10.1109/TPAMI.2010.86}
\showDOI{\tempurl}


\bibitem[\protect\citeauthoryear{Bulling, Weichel, and Gellersen}{Bulling
  et~al\mbox{.}}{2013}]%
        {bulling13_chi}
\bibfield{author}{\bibinfo{person}{Andreas Bulling}, \bibinfo{person}{Christian
  Weichel}, {and} \bibinfo{person}{Hans Gellersen}.}
  \bibinfo{year}{2013}\natexlab{}.
\newblock \showarticletitle{EyeContext: Recognition of High-level Contextual
  Cues from Human Visual Behaviour}. In \bibinfo{booktitle}{\emph{Proc. ACM
  SIGCHI Conference on Human Factors in Computing Systems (CHI)}}.
  \bibinfo{pages}{305--308}.
\newblock
\urldef\tempurl%
\url{https://doi.org/10.1145/2470654.2470697}
\showDOI{\tempurl}


\bibitem[\protect\citeauthoryear{Caine}{Caine}{2009}]%
        {Caine.2009}
\bibfield{author}{\bibinfo{person}{Kelly Caine}.}
  \bibinfo{year}{2009}\natexlab{}.
\newblock \bibinfo{booktitle}{\emph{Exploring everyday privacy behaviors and
  misclosures}}.
\newblock \bibinfo{publisher}{Georgia Institute of Technology}.
\newblock


\bibitem[\protect\citeauthoryear{Chan and Minamizawa}{Chan and
  Minamizawa}{2017}]%
        {Chan.2017}
\bibfield{author}{\bibinfo{person}{Liwei Chan} {and} \bibinfo{person}{Kouta
  Minamizawa}.} \bibinfo{year}{2017}\natexlab{}.
\newblock \showarticletitle{FrontFace: Facilitating Communication Between HMD
  Users and Outsiders Using Front-facing-screen HMDs}. In
  \bibinfo{booktitle}{\emph{Proceedings of the 19th International Conference on
  Human-Computer Interaction with Mobile Devices and Services}}
  \emph{(\bibinfo{series}{MobileHCI '17})}. \bibinfo{publisher}{ACM},
  \bibinfo{address}{New York, NY, USA}, Article \bibinfo{articleno}{22},
  \bibinfo{numpages}{5}~pages.
\newblock
\showISBNx{978-1-4503-5075-4}
\urldef\tempurl%
\url{https://doi.org/10.1145/3098279.3098548}
\showDOI{\tempurl}


\bibitem[\protect\citeauthoryear{Chowdhury, Ferdous, and Jose}{Chowdhury
  et~al\mbox{.}}{2016}]%
        {Chowdhury.2016a}
\bibfield{author}{\bibinfo{person}{Soumyadeb Chowdhury},
  \bibinfo{person}{Md~Sadek Ferdous}, {and} \bibinfo{person}{Joemon~M Jose}.}
  \bibinfo{year}{2016}\natexlab{}.
\newblock \showarticletitle{Exploring Lifelog Sharing and Privacy}. In
  \bibinfo{booktitle}{\emph{Proceedings of the 2016 ACM International Joint
  Conference on Pervasive and Ubiquitous Computing: Adjunct}}
  \emph{(\bibinfo{series}{UbiComp '16})}. \bibinfo{publisher}{ACM},
  \bibinfo{address}{New York, NY, USA}, \bibinfo{pages}{553--558}.
\newblock
\showISBNx{978-1-4503-4462-3}
\urldef\tempurl%
\url{https://doi.org/10.1145/2968219.2968320}
\showDOI{\tempurl}


\bibitem[\protect\citeauthoryear{Denning, Dehlawi, and Kohno}{Denning
  et~al\mbox{.}}{2014}]%
        {Denning.2014}
\bibfield{author}{\bibinfo{person}{Tamara Denning}, \bibinfo{person}{Zakariya
  Dehlawi}, {and} \bibinfo{person}{Tadayoshi Kohno}.}
  \bibinfo{year}{2014}\natexlab{}.
\newblock \showarticletitle{In situ with Bystanders of Augmented Reality
  Glasses: Perspectives on Recording and Privacy-mediating Technologies}. In
  \bibinfo{booktitle}{\emph{Proceedings of the Conference on Human Factors in
  Computing Systems (CHI)}}. ACM, \bibinfo{pages}{2377--2386}.
\newblock
\urldef\tempurl%
\url{https://doi.org/10.1145/2556288.2557352}
\showDOI{\tempurl}


\bibitem[\protect\citeauthoryear{Ens, Grossman, Anderson, Matejka, and
  Fitzmaurice}{Ens et~al\mbox{.}}{2015}]%
        {Ens.2015}
\bibfield{author}{\bibinfo{person}{Barrett Ens}, \bibinfo{person}{Tovi
  Grossman}, \bibinfo{person}{Fraser Anderson}, \bibinfo{person}{Justin
  Matejka}, {and} \bibinfo{person}{George Fitzmaurice}.}
  \bibinfo{year}{2015}\natexlab{}.
\newblock \showarticletitle{Candid interaction: Revealing hidden mobile and
  wearable computing activities}. In \bibinfo{booktitle}{\emph{Proceedings of
  the 28th Annual ACM Symposium on User Interface Software \& Technology}}.
  ACM, \bibinfo{pages}{467--476}.
\newblock
\urldef\tempurl%
\url{https://doi.org/10.1145/2807442.2807449}
\showDOI{\tempurl}


\bibitem[\protect\citeauthoryear{Erickson, Compiano, and Shin}{Erickson
  et~al\mbox{.}}{2014}]%
        {ericksonneural}
\bibfield{author}{\bibinfo{person}{Zackory Erickson}, \bibinfo{person}{Jared
  Compiano}, {and} \bibinfo{person}{Richard Shin}.}
  \bibinfo{year}{2014}\natexlab{}.
\newblock \showarticletitle{Neural Networks for Improving Wearable Device
  Security}.
\newblock  (\bibinfo{year}{2014}).
\newblock


\bibitem[\protect\citeauthoryear{Eriksen and Yeh}{Eriksen and Yeh}{1985}]%
        {eriksen1985allocation}
\bibfield{author}{\bibinfo{person}{Charles~W Eriksen} {and}
  \bibinfo{person}{Yei-yu Yeh}.} \bibinfo{year}{1985}\natexlab{}.
\newblock \showarticletitle{Allocation of attention in the visual field}.
\newblock \bibinfo{journal}{\emph{Journal of Experimental Psychology: Human
  Perception and Performance}} \bibinfo{volume}{11}, \bibinfo{number}{5}
  (\bibinfo{year}{1985}), \bibinfo{pages}{583}.
\newblock
\urldef\tempurl%
\url{https://doi.org/10.1037/0096-1523.11.5.583}
\showDOI{\tempurl}


\bibitem[\protect\citeauthoryear{Ferdous, Chowdhury, and Jose}{Ferdous
  et~al\mbox{.}}{2017}]%
        {Ferdous.2017}
\bibfield{author}{\bibinfo{person}{Md~Sadek Ferdous},
  \bibinfo{person}{Soumyadeb Chowdhury}, {and} \bibinfo{person}{Joemon~M
  Jose}.} \bibinfo{year}{2017}\natexlab{}.
\newblock \showarticletitle{Analysing privacy in visual lifelogging}.
\newblock \bibinfo{journal}{\emph{Pervasive and Mobile Computing}}
  (\bibinfo{year}{2017}).
\newblock
\urldef\tempurl%
\url{https://doi.org/10.1016/j.pmcj.2017.03.003}
\showDOI{\tempurl}


\bibitem[\protect\citeauthoryear{Fischer}{Fischer}{2001}]%
        {fischer2001user}
\bibfield{author}{\bibinfo{person}{Gerhard Fischer}.}
  \bibinfo{year}{2001}\natexlab{}.
\newblock \showarticletitle{User modeling in human--computer interaction}.
\newblock \bibinfo{journal}{\emph{User modeling and user-adapted interaction}}
  \bibinfo{volume}{11}, \bibinfo{number}{1-2} (\bibinfo{year}{2001}),
  \bibinfo{pages}{65--86}.
\newblock
\urldef\tempurl%
\url{https://doi.org/10.1023/A:1011145532042}
\showDOI{\tempurl}


\bibitem[\protect\citeauthoryear{H\"{a}kkil\"{a}, Vahabpour, Colley,
  V\"{a}yrynen, and Koskela}{H\"{a}kkil\"{a} et~al\mbox{.}}{2015}]%
        {Hakkila.2015}
\bibfield{author}{\bibinfo{person}{Jonna H\"{a}kkil\"{a}},
  \bibinfo{person}{Farnaz Vahabpour}, \bibinfo{person}{Ashley Colley},
  \bibinfo{person}{Jani V\"{a}yrynen}, {and} \bibinfo{person}{Timo Koskela}.}
  \bibinfo{year}{2015}\natexlab{}.
\newblock \showarticletitle{Design Probes Study on User Perceptions of a Smart
  Glasses Concept}. In \bibinfo{booktitle}{\emph{Proceedings of the 14th
  International Conference on Mobile and Ubiquitous Multimedia}}
  \emph{(\bibinfo{series}{MUM '15})}. \bibinfo{publisher}{ACM},
  \bibinfo{address}{New York, NY, USA}, \bibinfo{pages}{223--233}.
\newblock
\showISBNx{978-1-4503-3605-5}
\urldef\tempurl%
\url{https://doi.org/10.1145/2836041.2836064}
\showDOI{\tempurl}


\bibitem[\protect\citeauthoryear{Hansen, Johansen, Hansen, Itoh, and
  Mashino}{Hansen et~al\mbox{.}}{2003}]%
        {hansen2003command}
\bibfield{author}{\bibinfo{person}{John~Paulin Hansen},
  \bibinfo{person}{Anders~Sewerin Johansen}, \bibinfo{person}{Dan~Witzner
  Hansen}, \bibinfo{person}{Kenji Itoh}, {and} \bibinfo{person}{Satoru
  Mashino}.} \bibinfo{year}{2003}\natexlab{}.
\newblock \showarticletitle{Command without a click: Dwell time typing by mouse
  and gaze selections}. In \bibinfo{booktitle}{\emph{Proceedings of
  Human-Computer Interaction--INTERACT}}. \bibinfo{pages}{121--128}.
\newblock


\bibitem[\protect\citeauthoryear{Harvey}{Harvey}{2010}]%
        {Harvey.2010}
\bibfield{author}{\bibinfo{person}{Adam Harvey}.}
  \bibinfo{year}{2010}\natexlab{}.
\newblock \showarticletitle{Camoflash-anti-paparazzi clutch}.
\newblock  (\bibinfo{year}{2010}).
\newblock
\urldef\tempurl%
\url{http://ahprojects. com/projects/camoflash/}
\showURL{%
\tempurl}
\newblock
\shownote{accessed 13/09/2017.}


\bibitem[\protect\citeauthoryear{Harvey}{Harvey}{2012}]%
        {Harvey.2012}
\bibfield{author}{\bibinfo{person}{Adam Harvey}.}
  \bibinfo{year}{2012}\natexlab{}.
\newblock \showarticletitle{CVDazzle: Camouflage from Computer Vision}.
\newblock \bibinfo{journal}{\emph{Technical report}} (\bibinfo{year}{2012}).
\newblock


\bibitem[\protect\citeauthoryear{Hoppe, Loetscher, Morey, and Bulling}{Hoppe
  et~al\mbox{.}}{2018}]%
        {hoppe18_fhns}
\bibfield{author}{\bibinfo{person}{Sabrina Hoppe}, \bibinfo{person}{Tobias
  Loetscher}, \bibinfo{person}{Stephanie Morey}, {and} \bibinfo{person}{Andreas
  Bulling}.} \bibinfo{year}{2018}\natexlab{}.
\newblock \showarticletitle{Eye Movements During Everyday Behavior Predict
  Personality Traits}.
\newblock \bibinfo{journal}{\emph{Frontiers in Human Neuroscience}}
  \bibinfo{volume}{12} (\bibinfo{year}{2018}).
\newblock
\urldef\tempurl%
\url{https://doi.org/10.3389/fnhum.2018.00105}
\showDOI{\tempurl}


\bibitem[\protect\citeauthoryear{Hoyle, Templeman, Anthony, Crandall, and
  Kapadia}{Hoyle et~al\mbox{.}}{2015}]%
        {Hoyle.2015}
\bibfield{author}{\bibinfo{person}{Roberto Hoyle}, \bibinfo{person}{Robert
  Templeman}, \bibinfo{person}{Denise Anthony}, \bibinfo{person}{David
  Crandall}, {and} \bibinfo{person}{Apu Kapadia}.}
  \bibinfo{year}{2015}\natexlab{}.
\newblock \showarticletitle{Sensitive lifelogs: A privacy analysis of photos
  from wearable cameras}. In \bibinfo{booktitle}{\emph{Proceedings of the 33rd
  Annual ACM conference on human factors in computing systems}}. ACM,
  \bibinfo{pages}{1645--1648}.
\newblock
\urldef\tempurl%
\url{https://doi.org/10.1145/2702123.2702183}
\showDOI{\tempurl}


\bibitem[\protect\citeauthoryear{Hoyle, Templeman, Armes, Anthony, Crandall,
  and Kapadia}{Hoyle et~al\mbox{.}}{2014}]%
        {Hoyle.2014}
\bibfield{author}{\bibinfo{person}{Roberto Hoyle}, \bibinfo{person}{Robert
  Templeman}, \bibinfo{person}{Steven Armes}, \bibinfo{person}{Denise Anthony},
  \bibinfo{person}{David Crandall}, {and} \bibinfo{person}{Apu Kapadia}.}
  \bibinfo{year}{2014}\natexlab{}.
\newblock \showarticletitle{Privacy Behaviors of Lifeloggers using Wearable
  Cameras}. In \bibinfo{booktitle}{\emph{International Joint Conference on
  Pervasive and Ubiquitous Computing (Ubicomp)}}. ACM,
  \bibinfo{pages}{571--582}.
\newblock
\urldef\tempurl%
\url{https://doi.org/10.1145/2632048.2632079}
\showDOI{\tempurl}


\bibitem[\protect\citeauthoryear{Itti and Koch}{Itti and Koch}{2001}]%
        {itti2001computational}
\bibfield{author}{\bibinfo{person}{Laurent Itti} {and}
  \bibinfo{person}{Christof Koch}.} \bibinfo{year}{2001}\natexlab{}.
\newblock \showarticletitle{Computational modelling of visual attention}.
\newblock \bibinfo{journal}{\emph{Nature reviews neuroscience}}
  \bibinfo{volume}{2}, \bibinfo{number}{3} (\bibinfo{year}{2001}),
  \bibinfo{pages}{194}.
\newblock
\urldef\tempurl%
\url{https://doi.org/10.1038/35058500}
\showDOI{\tempurl}


\bibitem[\protect\citeauthoryear{Kassner, Patera, and Bulling}{Kassner
  et~al\mbox{.}}{2014}]%
        {Kassner.2014}
\bibfield{author}{\bibinfo{person}{Moritz Kassner}, \bibinfo{person}{William
  Patera}, {and} \bibinfo{person}{Andreas Bulling}.}
  \bibinfo{year}{2014}\natexlab{}.
\newblock \showarticletitle{Pupil: an open source platform for pervasive eye
  tracking and mobile gaze-based interaction}. In
  \bibinfo{booktitle}{\emph{Adj. Proc. UbiComp}}. \bibinfo{pages}{1151--1160}.
\newblock
\showISBNx{978-1-4503-3047-3}
\urldef\tempurl%
\url{http://dx.doi.org/10.1145/2638728.2641695}
\showURL{%
\tempurl}


\bibitem[\protect\citeauthoryear{Koelle, Heuten, and Boll}{Koelle
  et~al\mbox{.}}{2017}]%
        {Koelle.2017}
\bibfield{author}{\bibinfo{person}{Marion Koelle}, \bibinfo{person}{Wilko
  Heuten}, {and} \bibinfo{person}{Susanne Boll}.}
  \bibinfo{year}{2017}\natexlab{}.
\newblock \showarticletitle{Are You Hiding It?: Usage Habits of Lifelogging
  Camera Wearers}. In \bibinfo{booktitle}{\emph{Proceedings of the 19th
  International Conference on Human-Computer Interaction with Mobile Devices
  and Services}} \emph{(\bibinfo{series}{MobileHCI '17})}.
  \bibinfo{publisher}{ACM}, \bibinfo{address}{New York, NY, USA}, Article
  \bibinfo{articleno}{80}, \bibinfo{numpages}{8}~pages.
\newblock
\showISBNx{978-1-4503-5075-4}
\urldef\tempurl%
\url{https://doi.org/10.1145/3098279.3122123}
\showDOI{\tempurl}


\bibitem[\protect\citeauthoryear{Koelle, Kranz, and M{\"o}ller}{Koelle
  et~al\mbox{.}}{2015}]%
        {Koelle.2015}
\bibfield{author}{\bibinfo{person}{Marion Koelle}, \bibinfo{person}{Matthias
  Kranz}, {and} \bibinfo{person}{Andreas M{\"o}ller}.}
  \bibinfo{year}{2015}\natexlab{}.
\newblock \showarticletitle{Don't look at me that way!: Understanding User
  Attitudes Towards Data Glasses Usage}. In
  \bibinfo{booktitle}{\emph{Proceedings of the 17th international conference on
  human-computer interaction with mobile devices and services}}. ACM,
  \bibinfo{pages}{362--372}.
\newblock
\urldef\tempurl%
\url{https://doi.org/10.1145/2785830.2785842}
\showDOI{\tempurl}


\bibitem[\protect\citeauthoryear{Koelle, Wolf, and Boll}{Koelle
  et~al\mbox{.}}{2018}]%
        {koelle2018beyond}
\bibfield{author}{\bibinfo{person}{Marion Koelle}, \bibinfo{person}{Katrin
  Wolf}, {and} \bibinfo{person}{Susanne Boll}.}
  \bibinfo{year}{2018}\natexlab{}.
\newblock \showarticletitle{Beyond LED Status Lights-Design Requirements of
  Privacy Notices for Body-worn Cameras}. In
  \bibinfo{booktitle}{\emph{Proceedings of the Twelfth International Conference
  on Tangible, Embedded, and Embodied Interaction}}. ACM,
  \bibinfo{pages}{177--187}.
\newblock
\urldef\tempurl%
\url{https://doi.org/10.1145/3173225.3173234}
\showDOI{\tempurl}


\bibitem[\protect\citeauthoryear{Korayem, Templeman, Chen, Crandall, and
  Kapadia}{Korayem et~al\mbox{.}}{2014}]%
        {Korayem.2014}
\bibfield{author}{\bibinfo{person}{Mohammed Korayem}, \bibinfo{person}{Robert
  Templeman}, \bibinfo{person}{Dennis Chen}, \bibinfo{person}{David Crandall},
  {and} \bibinfo{person}{Apu Kapadia}.} \bibinfo{year}{2014}\natexlab{}.
\newblock \showarticletitle{Screenavoider: Protecting computer screens from
  ubiquitous cameras}.
\newblock \bibinfo{journal}{\emph{arXiv preprint arXiv:1412.0008}}
  (\bibinfo{year}{2014}).
\newblock


\bibitem[\protect\citeauthoryear{Korayem, Templeman, Chen, Crandall, and
  Kapadia}{Korayem et~al\mbox{.}}{2016}]%
        {korayem2016enhancing}
\bibfield{author}{\bibinfo{person}{Mohammed Korayem}, \bibinfo{person}{Robert
  Templeman}, \bibinfo{person}{Dennis Chen}, \bibinfo{person}{David Crandall},
  {and} \bibinfo{person}{Apu Kapadia}.} \bibinfo{year}{2016}\natexlab{}.
\newblock \showarticletitle{Enhancing lifelogging privacy by detecting
  screens}. In \bibinfo{booktitle}{\emph{Proceedings of the 2016 CHI Conference
  on Human Factors in Computing Systems}}. ACM, \bibinfo{pages}{4309--4314}.
\newblock
\urldef\tempurl%
\url{https://doi.org/10.1145/2702123.2702183}
\showDOI{\tempurl}


\bibitem[\protect\citeauthoryear{Krombholz, Dabrowski, Smith, and
  Weippl}{Krombholz et~al\mbox{.}}{2015}]%
        {Krombholz.2015}
\bibfield{author}{\bibinfo{person}{Katharina Krombholz},
  \bibinfo{person}{Adrian Dabrowski}, \bibinfo{person}{Matthew Smith}, {and}
  \bibinfo{person}{Edgar Weippl}.} \bibinfo{year}{2015}\natexlab{}.
\newblock \showarticletitle{Ok glass, leave me alone: towards a systematization
  of privacy enhancing technologies for wearable computing}. In
  \bibinfo{booktitle}{\emph{International Conference on Financial Cryptography
  and Data Security}}. Springer, \bibinfo{pages}{274--280}.
\newblock
\urldef\tempurl%
\url{https://doi.org/10.1007/978-3-662-48051-9_20}
\showDOI{\tempurl}


\bibitem[\protect\citeauthoryear{Liebling and Preibusch}{Liebling and
  Preibusch}{2014}]%
        {liebling2014privacy}
\bibfield{author}{\bibinfo{person}{Daniel~J Liebling} {and}
  \bibinfo{person}{S{\"o}ren Preibusch}.} \bibinfo{year}{2014}\natexlab{}.
\newblock \showarticletitle{Privacy considerations for a pervasive eye tracking
  world}. In \bibinfo{booktitle}{\emph{Proceedings of the 2014 ACM
  International Joint Conference on Pervasive and Ubiquitous Computing: Adjunct
  Publication}}. ACM, \bibinfo{pages}{1169--1177}.
\newblock
\urldef\tempurl%
\url{https://doi.org/10.1145/2638728.2641688}
\showDOI{\tempurl}


\bibitem[\protect\citeauthoryear{Mayring}{Mayring}{2014}]%
        {Mayring.2014}
\bibfield{author}{\bibinfo{person}{Philipp Mayring}.}
  \bibinfo{year}{2014}\natexlab{}.
\newblock \bibinfo{booktitle}{\emph{Qualitative content analysis: theoretical
  foundation, basic procedures and software solution}}.
\newblock 143 pages.
\newblock


\bibitem[\protect\citeauthoryear{Motti and Caine}{Motti and Caine}{2016}]%
        {Motti.2016}
\bibfield{author}{\bibinfo{person}{Vivian~Genaro Motti} {and}
  \bibinfo{person}{Kelly Caine}.} \bibinfo{year}{2016}\natexlab{}.
\newblock \showarticletitle{Towards a Visual Vocabulary for Privacy Concepts}.
  In \bibinfo{booktitle}{\emph{Proceedings of the Human Factors and Ergonomics
  Society Annual Meeting}}, Vol.~\bibinfo{volume}{60}. SAGE Publications Sage
  CA: Los Angeles, CA, \bibinfo{pages}{1078--1082}.
\newblock
\urldef\tempurl%
\url{https://doi.org/10.1177/1541931213601249}
\showDOI{\tempurl}


\bibitem[\protect\citeauthoryear{Orekondy, Schiele, and Fritz}{Orekondy
  et~al\mbox{.}}{2017}]%
        {orekondy17iccv}
\bibfield{author}{\bibinfo{person}{Tribhuvanesh Orekondy},
  \bibinfo{person}{Bernt Schiele}, {and} \bibinfo{person}{Mario Fritz}.}
  \bibinfo{year}{2017}\natexlab{}.
\newblock \showarticletitle{Towards a Visual Privacy Advisor: Understanding and
  Predicting Privacy Risks in Images}. In
  \bibinfo{booktitle}{\emph{International Conference on Computer Vision (ICCV
  2017)}}. \bibinfo{address}{Venice, Italy}.
\newblock
\urldef\tempurl%
\url{https://doi.org/10.1109/ICCV.2017.398}
\showDOI{\tempurl}


\bibitem[\protect\citeauthoryear{Perez, Zeadally, and Griffith}{Perez
  et~al\mbox{.}}{2017}]%
        {perez2017bystanders}
\bibfield{author}{\bibinfo{person}{Alfredo~J Perez}, \bibinfo{person}{Sherali
  Zeadally}, {and} \bibinfo{person}{Scott Griffith}.}
  \bibinfo{year}{2017}\natexlab{}.
\newblock \showarticletitle{Bystanders' Privacy}.
\newblock \bibinfo{journal}{\emph{IT Professional}} \bibinfo{volume}{19},
  \bibinfo{number}{3} (\bibinfo{year}{2017}), \bibinfo{pages}{61--65}.
\newblock
\urldef\tempurl%
\url{https://doi.org/10.1109/MITP.2017.42}
\showDOI{\tempurl}


\bibitem[\protect\citeauthoryear{Portnoff, Lee, Egelman, Mishra, Leung, and
  Wagner}{Portnoff et~al\mbox{.}}{2015}]%
        {Portnoff.2015}
\bibfield{author}{\bibinfo{person}{Rebecca~S Portnoff},
  \bibinfo{person}{Linda~N Lee}, \bibinfo{person}{Serge Egelman},
  \bibinfo{person}{Pratyush Mishra}, \bibinfo{person}{Derek Leung}, {and}
  \bibinfo{person}{David Wagner}.} \bibinfo{year}{2015}\natexlab{}.
\newblock \showarticletitle{Somebody's watching me?: Assessing the
  effectiveness of Webcam indicator lights}. In
  \bibinfo{booktitle}{\emph{Proceedings of the 33rd Annual ACM Conference on
  Human Factors in Computing Systems}}. ACM, \bibinfo{pages}{1649--1658}.
\newblock
\urldef\tempurl%
\url{https://doi.org/10.1145/2702123.2702611}
\showDOI{\tempurl}


\bibitem[\protect\citeauthoryear{Pousman, Iachello, Fithian, Moghazy, and
  Stasko}{Pousman et~al\mbox{.}}{2004}]%
        {Pousman.2004}
\bibfield{author}{\bibinfo{person}{Zachary Pousman}, \bibinfo{person}{Giovanni
  Iachello}, \bibinfo{person}{Rachel Fithian}, \bibinfo{person}{Jehan Moghazy},
  {and} \bibinfo{person}{John Stasko}.} \bibinfo{year}{2004}\natexlab{}.
\newblock \showarticletitle{Design iterations for a location-aware event
  planner}.
\newblock \bibinfo{journal}{\emph{Personal and Ubiquitous Computing}}
  \bibinfo{volume}{8}, \bibinfo{number}{2} (\bibinfo{year}{2004}),
  \bibinfo{pages}{117--125}.
\newblock
\urldef\tempurl%
\url{https://doi.org/10.1007/s00779-004-0266-y}
\showDOI{\tempurl}


\bibitem[\protect\citeauthoryear{Preibusch}{Preibusch}{2014}]%
        {Preibusch.2014}
\bibfield{author}{\bibinfo{person}{S\"{o}ren Preibusch}.}
  \bibinfo{year}{2014}\natexlab{}.
\newblock \showarticletitle{Eye-tracking. Privacy interfaces for the next
  ubiquitous modality}. In \bibinfo{booktitle}{\emph{2014 W3C Workshop on
  Privacy and User-Centric Controls}}.
\newblock
\urldef\tempurl%
\url{https://www.w3.org/2014/privacyws/pp/Preibusch.pdf}
\showURL{%
\tempurl}


\bibitem[\protect\citeauthoryear{Price, Stuart, Calikli, Mccormick, Mehta,
  Hutton, Bandara, Levine, and Nuseibeh}{Price et~al\mbox{.}}{2017}]%
        {Price.2017}
\bibfield{author}{\bibinfo{person}{Blaine~A. Price}, \bibinfo{person}{Avelie
  Stuart}, \bibinfo{person}{Gul Calikli}, \bibinfo{person}{Ciaran Mccormick},
  \bibinfo{person}{Vikram Mehta}, \bibinfo{person}{Luke Hutton},
  \bibinfo{person}{Arosha~K. Bandara}, \bibinfo{person}{Mark Levine}, {and}
  \bibinfo{person}{Bashar Nuseibeh}.} \bibinfo{year}{2017}\natexlab{}.
\newblock \showarticletitle{Logging You, Logging Me: A Replicable Study of
  Privacy and Sharing Behaviour in Groups of Visual Lifeloggers}.
\newblock \bibinfo{journal}{\emph{Proc. ACM Interact. Mob. Wearable Ubiquitous
  Technol.}} \bibinfo{volume}{1}, \bibinfo{number}{2}, Article
  \bibinfo{articleno}{22} (\bibinfo{date}{June} \bibinfo{year}{2017}),
  \bibinfo{numpages}{18}~pages.
\newblock
\showISSN{2474-9567}
\urldef\tempurl%
\url{https://doi.org/10.1145/3090087}
\showDOI{\tempurl}


\bibitem[\protect\citeauthoryear{Profita, Albaghli, Findlater, Jaeger, and
  Kane}{Profita et~al\mbox{.}}{2016}]%
        {Profita.2016}
\bibfield{author}{\bibinfo{person}{Halley Profita}, \bibinfo{person}{Reem
  Albaghli}, \bibinfo{person}{Leah Findlater}, \bibinfo{person}{Paul Jaeger},
  {and} \bibinfo{person}{Shaun~K Kane}.} \bibinfo{year}{2016}\natexlab{}.
\newblock \showarticletitle{The AT Effect: How Disability Affects the Perceived
  Social Acceptability of Head-Mounted Display Use}. In
  \bibinfo{booktitle}{\emph{Proceedings of the 2016 CHI Conference on Human
  Factors in Computing Systems}}. ACM, \bibinfo{pages}{4884--4895}.
\newblock
\urldef\tempurl%
\url{https://doi.org/10.1145/2858036.2858130}
\showDOI{\tempurl}


\bibitem[\protect\citeauthoryear{Raval, Srivastava, Lebeck, Cox, and
  Machanavajjhala}{Raval et~al\mbox{.}}{2014}]%
        {Raval.2014}
\bibfield{author}{\bibinfo{person}{Nisarg Raval}, \bibinfo{person}{Animesh
  Srivastava}, \bibinfo{person}{Kiron Lebeck}, \bibinfo{person}{Landon Cox},
  {and} \bibinfo{person}{Ashwin Machanavajjhala}.}
  \bibinfo{year}{2014}\natexlab{}.
\newblock \showarticletitle{Markit: Privacy markers for protecting visual
  secrets}. In \bibinfo{booktitle}{\emph{Proceedings of the 2014 ACM
  International Joint Conference on Pervasive and Ubiquitous Computing: Adjunct
  Publication}}. ACM, \bibinfo{pages}{1289--1295}.
\newblock
\urldef\tempurl%
\url{https://doi.org/10.1145/2638728.2641707}
\showDOI{\tempurl}


\bibitem[\protect\citeauthoryear{Schiff, Meingast, Mulligan, Sastry, and
  Goldberg}{Schiff et~al\mbox{.}}{2007}]%
        {Schiff.2007}
\bibfield{author}{\bibinfo{person}{Jeremy Schiff}, \bibinfo{person}{Marci
  Meingast}, \bibinfo{person}{Deirdre~K Mulligan}, \bibinfo{person}{Shankar
  Sastry}, {and} \bibinfo{person}{Ken Goldberg}.}
  \bibinfo{year}{2007}\natexlab{}.
\newblock \showarticletitle{Respectful cameras: Detecting visual markers in
  real-time to address privacy concerns}. In
  \bibinfo{booktitle}{\emph{Intelligent Robots and Systems, 2007. IROS 2007.
  IEEE/RSJ International Conference on}}. IEEE, \bibinfo{pages}{971--978}.
\newblock
\urldef\tempurl%
\url{https://doi.org/10.1007/978-1-84882-301-3_5}
\showDOI{\tempurl}


\bibitem[\protect\citeauthoryear{Schlegel, Kapadia, and Lee}{Schlegel
  et~al\mbox{.}}{2011}]%
        {Schlegel.2011}
\bibfield{author}{\bibinfo{person}{Roman Schlegel}, \bibinfo{person}{Apu
  Kapadia}, {and} \bibinfo{person}{Adam~J Lee}.}
  \bibinfo{year}{2011}\natexlab{}.
\newblock \showarticletitle{Eyeing your exposure: quantifying and controlling
  information sharing for improved privacy}. In
  \bibinfo{booktitle}{\emph{Proceedings of the Seventh Symposium on Usable
  Privacy and Security}}. ACM, \bibinfo{pages}{14}.
\newblock
\urldef\tempurl%
\url{https://doi.org/10.1145/2078827.2078846}
\showDOI{\tempurl}


\bibitem[\protect\citeauthoryear{Shu, Zheng, and Hui}{Shu
  et~al\mbox{.}}{2016}]%
        {Shu.2016}
\bibfield{author}{\bibinfo{person}{Jiayu Shu}, \bibinfo{person}{Rui Zheng},
  {and} \bibinfo{person}{Pan Hui}.} \bibinfo{year}{2016}\natexlab{}.
\newblock \showarticletitle{Cardea: Context-aware visual privacy protection
  from pervasive cameras}.
\newblock \bibinfo{journal}{\emph{arXiv preprint arXiv:1610.00889}}
  (\bibinfo{year}{2016}).
\newblock


\bibitem[\protect\citeauthoryear{Steil and Bulling}{Steil and Bulling}{2015}]%
        {steil15_ubicomp}
\bibfield{author}{\bibinfo{person}{Julian Steil} {and} \bibinfo{person}{Andreas
  Bulling}.} \bibinfo{year}{2015}\natexlab{}.
\newblock \showarticletitle{Discovery of Everyday Human Activities From
  Long-term Visual Behaviour Using Topic Models}. In
  \bibinfo{booktitle}{\emph{Proc. ACM International Joint Conference on
  Pervasive and Ubiquitous Computing (UbiComp)}}. \bibinfo{pages}{75--85}.
\newblock
\urldef\tempurl%
\url{https://doi.org/10.1145/2750858.2807520}
\showDOI{\tempurl}


\bibitem[\protect\citeauthoryear{Steil, M{\"u}ller, Sugano, and Bulling}{Steil
  et~al\mbox{.}}{2018}]%
        {steil2018forecasting}
\bibfield{author}{\bibinfo{person}{Julian Steil}, \bibinfo{person}{Philipp
  M{\"u}ller}, \bibinfo{person}{Yusuke Sugano}, {and} \bibinfo{person}{Andreas
  Bulling}.} \bibinfo{year}{2018}\natexlab{}.
\newblock \showarticletitle{Forecasting user attention during everyday mobile
  interactions using device-integrated and wearable sensors}. In
  \bibinfo{booktitle}{\emph{Proceedings of the 20th International Conference on
  Human-Computer Interaction with Mobile Devices and Services}}. ACM,
  \bibinfo{pages}{1}.
\newblock
\urldef\tempurl%
\url{https://doi.org/10.1145/3229434.3229439}
\showDOI{\tempurl}


\bibitem[\protect\citeauthoryear{Sugano and Bulling}{Sugano and
  Bulling}{2015}]%
        {Sugano_UIST15}
\bibfield{author}{\bibinfo{person}{Yusuke Sugano} {and}
  \bibinfo{person}{Andreas Bulling}.} \bibinfo{year}{2015}\natexlab{}.
\newblock \showarticletitle{Self-Calibrating Head-Mounted Eye Trackers Using
  Egocentric Visual Saliency}. In \bibinfo{booktitle}{\emph{Proc. of the 28th
  ACM Symposium on User Interface Software and Technology (UIST 2015)}}.
  \bibinfo{pages}{363--372}.
\newblock
\urldef\tempurl%
\url{https://doi.org/10.1145/2807442.2807445}
\showDOI{\tempurl}


\bibitem[\protect\citeauthoryear{Sugano, Zhang, and Bulling}{Sugano
  et~al\mbox{.}}{2016}]%
        {sugano2016aggregaze}
\bibfield{author}{\bibinfo{person}{Yusuke Sugano}, \bibinfo{person}{Xucong
  Zhang}, {and} \bibinfo{person}{Andreas Bulling}.}
  \bibinfo{year}{2016}\natexlab{}.
\newblock \showarticletitle{Aggregaze: Collective estimation of audience
  attention on public displays}. In \bibinfo{booktitle}{\emph{Proceedings of
  the 29th Annual Symposium on User Interface Software and Technology}}. ACM,
  \bibinfo{pages}{821--831}.
\newblock
\urldef\tempurl%
\url{https://doi.org/10.1145/2984511.2984536}
\showDOI{\tempurl}


\bibitem[\protect\citeauthoryear{Szegedy, Liu, Jia, Sermanet, Reed, Anguelov,
  Erhan, Vanhoucke, and Rabinovich}{Szegedy et~al\mbox{.}}{2015}]%
        {43022}
\bibfield{author}{\bibinfo{person}{Christian Szegedy}, \bibinfo{person}{Wei
  Liu}, \bibinfo{person}{Yangqing Jia}, \bibinfo{person}{Pierre Sermanet},
  \bibinfo{person}{Scott Reed}, \bibinfo{person}{Dragomir Anguelov},
  \bibinfo{person}{Dumitru Erhan}, \bibinfo{person}{Vincent Vanhoucke}, {and}
  \bibinfo{person}{Andrew Rabinovich}.} \bibinfo{year}{2015}\natexlab{}.
\newblock \showarticletitle{Going Deeper with Convolutions}. In
  \bibinfo{booktitle}{\emph{Computer Vision and Pattern Recognition (CVPR)}}.
\newblock
\urldef\tempurl%
\url{http://arxiv.org/abs/1409.4842}
\showURL{%
\tempurl}


\bibitem[\protect\citeauthoryear{Templeman, Korayem, Crandall, and
  Kapadia}{Templeman et~al\mbox{.}}{2014}]%
        {Templeman.2014}
\bibfield{author}{\bibinfo{person}{Robert Templeman}, \bibinfo{person}{Mohammed
  Korayem}, \bibinfo{person}{David~J Crandall}, {and} \bibinfo{person}{Apu
  Kapadia}.} \bibinfo{year}{2014}\natexlab{}.
\newblock \showarticletitle{PlaceAvoider: Steering First-Person Cameras away
  from Sensitive Spaces}. In \bibinfo{booktitle}{\emph{NDSS}}.
\newblock
\urldef\tempurl%
\url{https://doi.org/10.14722/ndss.2014.23014}
\showDOI{\tempurl}


\bibitem[\protect\citeauthoryear{Tonsen, Steil, Sugano, and Bulling}{Tonsen
  et~al\mbox{.}}{2017}]%
        {tonsen17_imwut}
\bibfield{author}{\bibinfo{person}{Marc Tonsen}, \bibinfo{person}{Julian
  Steil}, \bibinfo{person}{Yusuke Sugano}, {and} \bibinfo{person}{Andreas
  Bulling}.} \bibinfo{year}{2017}\natexlab{}.
\newblock \showarticletitle{InvisibleEye: Mobile Eye Tracking Using Multiple
  Low-Resolution Cameras and Learning-Based Gaze Estimation}.
\newblock \bibinfo{journal}{\emph{Proceedings of the ACM on Interactive,
  Mobile, Wearable and Ubiquitous Technologies (IMWUT)}} \bibinfo{volume}{1},
  \bibinfo{number}{3} (\bibinfo{year}{2017}), \bibinfo{pages}{106:1--106:21}.
\newblock
\urldef\tempurl%
\url{https://doi.org/10.1145/3130971}
\showDOI{\tempurl}


\bibitem[\protect\citeauthoryear{Truong, Patel, Summet, and Abowd}{Truong
  et~al\mbox{.}}{2005}]%
        {Truong.2005}
\bibfield{author}{\bibinfo{person}{Khai Truong}, \bibinfo{person}{Shwetak
  Patel}, \bibinfo{person}{Jay Summet}, {and} \bibinfo{person}{Gregory Abowd}.}
  \bibinfo{year}{2005}\natexlab{}.
\newblock \showarticletitle{Preventing camera recording by designing a
  capture-resistant environment}.
\newblock \bibinfo{journal}{\emph{UbiComp 2005: Ubiquitous Computing}}
  (\bibinfo{year}{2005}), \bibinfo{pages}{903--903}.
\newblock
\urldef\tempurl%
\url{https://doi.org/10.1007/11551201_5}
\showDOI{\tempurl}


\bibitem[\protect\citeauthoryear{Vertegaal et~al\mbox{.}}{Vertegaal
  et~al\mbox{.}}{2003}]%
        {vertegaal2003attentive}
\bibfield{author}{\bibinfo{person}{Roel Vertegaal} {et~al\mbox{.}}}
  \bibinfo{year}{2003}\natexlab{}.
\newblock \showarticletitle{Attentive user interfaces}.
\newblock \bibinfo{journal}{\emph{Commun. ACM}} \bibinfo{volume}{46},
  \bibinfo{number}{3} (\bibinfo{year}{2003}), \bibinfo{pages}{30--33}.
\newblock
\urldef\tempurl%
\url{https://doi.org/10.1145/636772.636794}
\showDOI{\tempurl}


\bibitem[\protect\citeauthoryear{Westin}{Westin}{2003}]%
        {Westin.2003}
\bibfield{author}{\bibinfo{person}{Alan~F Westin}.}
  \bibinfo{year}{2003}\natexlab{}.
\newblock \showarticletitle{Social and political dimensions of privacy}.
\newblock \bibinfo{journal}{\emph{Journal of social issues}}
  \bibinfo{volume}{59}, \bibinfo{number}{2} (\bibinfo{year}{2003}),
  \bibinfo{pages}{431--453}.
\newblock
\urldef\tempurl%
\url{https://doi.org/10.1111/1540-4560.00072}
\showDOI{\tempurl}


\bibitem[\protect\citeauthoryear{Yamada, Gohshi, and Echizen}{Yamada
  et~al\mbox{.}}{2013}]%
        {yamada2013privacy}
\bibfield{author}{\bibinfo{person}{Takayuki Yamada}, \bibinfo{person}{Seiichi
  Gohshi}, {and} \bibinfo{person}{Isao Echizen}.}
  \bibinfo{year}{2013}\natexlab{}.
\newblock \showarticletitle{Privacy visor: Method based on light absorbing and
  reflecting properties for preventing face image detection}. In
  \bibinfo{booktitle}{\emph{Systems, Man, and Cybernetics (SMC), 2013 IEEE
  International Conference on}}. IEEE, \bibinfo{pages}{1572--1577}.
\newblock
\urldef\tempurl%
\url{https://doi.org/10.1109/SMC.2013.271}
\showDOI{\tempurl}


\bibitem[\protect\citeauthoryear{Yu, Zhang, Kuang, Lin, and Fan}{Yu
  et~al\mbox{.}}{2017}]%
        {yu2017iprivacy}
\bibfield{author}{\bibinfo{person}{Jun Yu}, \bibinfo{person}{Baopeng Zhang},
  \bibinfo{person}{Zhengzhong Kuang}, \bibinfo{person}{Dan Lin}, {and}
  \bibinfo{person}{Jianping Fan}.} \bibinfo{year}{2017}\natexlab{}.
\newblock \showarticletitle{iPrivacy: image privacy protection by identifying
  sensitive objects via deep multi-task learning}.
\newblock \bibinfo{journal}{\emph{IEEE Transactions on Information Forensics
  and Security}} \bibinfo{volume}{12}, \bibinfo{number}{5}
  (\bibinfo{year}{2017}), \bibinfo{pages}{1005--1016}.
\newblock
\urldef\tempurl%
\url{https://doi.org/10.1109/TIFS.2016.2636090}
\showDOI{\tempurl}


\end{thebibliography}
